\documentclass[aos, preprint]{imsart}

\RequirePackage[OT1]{fontenc}
\RequirePackage{amsthm,amsmath,natbib}

\usepackage{array}
\usepackage{mathrsfs}
\usepackage{amssymb,bbm}
\usepackage{amsfonts}
\usepackage{amsmath,amssymb}
\usepackage{graphicx}
\usepackage{amscd}
\usepackage{color}
\usepackage{float}
\usepackage{mathrsfs}
\usepackage{latexsym,bm}
\usepackage{array}
\usepackage{threeparttable}
\usepackage{enumerate}
\usepackage{rotating}
\usepackage{booktabs}
\usepackage{rotating}
\usepackage{multirow}
\usepackage{subfigure}
\usepackage[plain,noend]{algorithm2e}

\usepackage{geometry}
\DeclareFontFamily{U}{mathx}{\hyphenchar\font45}
\DeclareFontShape{U}{mathx}{m}{n}{
      <5> <6> <7> <8> <9> <10>
      <10.95> <12> <14.4> <17.28> <20.74> <24.88>
      mathx10
      }{}
\DeclareSymbolFont{mathx}{U}{mathx}{m}{n}
\DeclareMathAccent{\widecheck}{\mathalpha}{mathx}{"71}

\DeclareMathAccent{\widecheck}{\mathalpha}{mathx}{"71}

\startlocaldefs \numberwithin{equation}{section}
\theoremstyle{it}
\newtheorem{thm}{Theorem}[section]
\newtheorem{lemma}{Lemma}[section]
\newtheorem{ass}{Assumption}[section]
\newenvironment{thmbis}[1]
{\renewcommand{\theass}{\ref{#1}$'$}%
	\addtocounter{ass}{-1}%
	\begin{ass}}
	{\end{ass}}

\newtheorem{alg}{Algorithm}[section]
\endlocaldefs

\begin{document}

\begin{frontmatter}
\title{Non-standard inference for augmented double autoregressive models with null volatility coefficients}
\runtitle{Inference for augmented DAR models}

\begin{aug}
\author{\fnms{Feiyu} \snm{Jiang}\thanksref{m1}\ead[label=e1]{jfy16@mails.tsinghua.edu.cn}},
\author{\fnms{Dong} \snm{Li}\thanksref{m2}\ead[label=e2]{malidong@tsinghua.edu.cn}}
\and
\author{\fnms{Ke} \snm{Zhu}\thanksref{m3}\ead[label=e3]{mazhuke@hku.hk}
\ead[label=u1, url]{http://www.foo.com}}
\runauthor{Jiang et al.}

\affiliation{Tsinghua University\thanksmark{m1}\thanksmark{m2} and University of Hong Kong\thanksmark{m3}}

\address{
Center for Statistical Science\\ \quad and Department of Industry Engineering\\
Tsinghua University\\
Beijing 100084, China\\
\printead{e1}\\
\phantom{E-mail:\ }\printead*{e2}
}

\address
{Department of Statistics \& Actuarial Science\\
The University of Hong Kong\\
Hong Kong\\
\printead{e3}
}
\end{aug}

\begin{abstract}
This paper considers an augmented double autoregressive (DAR) model, which allows null volatility
coefficients to circumvent the over-parameterization problem in the  DAR model.
Since the volatility coefficients might be on the boundary,
the statistical inference methods based on the Gaussian quasi-maximum likelihood estimation (GQMLE)
become non-standard, and their asymptotics require the data to have a finite sixth moment,
which narrows  applicable scope in studying heavy-tailed data.
 To overcome this deficiency, this paper develops a systematic statistical inference procedure  based on the self-weighted GQMLE for the augmented DAR model. Except for the Lagrange multiplier test statistic,  the Wald, quasi-likelihood ratio and portmanteau test statistics are all shown to have non-standard asymptotics.
 The entire procedure is valid as long as  the data is stationary, and its usefulness is illustrated by simulation studies and one real example.
\end{abstract}



\end{frontmatter}

\section{Introduction}
Modelling conditional mean and volatility dynamics together is of extreme importance in econometrics and finance. A myriad of specifications have been proposed for the purpose, and among them, the double autoregressive (DAR) model has recently been attracting much attention in the literature, and it is defined as
\begin{flalign}\label{dar_model}
y_t=u+\sum_{i=1}^{p}\phi_{i}y_{t-i}+\eta_t\sqrt{\omega+\sum_{i=1}^{p}\alpha_{i}y_{t-i}^2},
\quad t=0, \pm1, ...,
\end{flalign}
where $u, \phi_{i}\in \mathbb{R}$,  $\omega>0, \alpha_{i}> 0$,
$\{\eta_t\}$ is a sequence of independent and identically distributed (i.i.d.)  random variables
with zero mean and unit variance,
and $\eta_t$ is independent of $\{y_s; s<t\}$. Model (\ref{dar_model}) was first termed by
\cite{ling2004estimation},
and it is a subclass of
ARMA-ARCH models in  \cite{weiss1986asymptotic} and of nonlinear AR models in \cite{cline2004stability}, but it is different from Engle's ARCH model if some $\phi_{i}\neq 0$.

As was shown in \cite{ling2007double}, model (\ref{dar_model}) has an important feature that its Gaussian quasi-maximum likelihood estimator (GQMLE) is asymptotically normal as long as $y_{t}$ has a finite fractional moment, while the ARMA-GARCH model (see, e.g., \cite{ling2007self} and \cite{zhangling2015}) does not.
This feature makes model (\ref{dar_model}) feasible and convenient to fit the often observed heavy-tailed data in applications,
but it relies on a crucial assumption that each volatility coefficient $\alpha_{i}$ has a positive lower bound,
which might result in the over-parameterization problem.
Moreover, both the conditional mean and volatility specifications in model (\ref{dar_model}) have the same order $p$.
This could be another shortcoming of model (\ref{dar_model}) and narrow down its applications.
Motivated by these facts, this paper considers an augmented DAR (ADAR) model of order $(p, q)$:
\begin{flalign}\label{adar_model}
y_t=u+\sum_{i=1}^{p}\phi_{i}y_{t-i}+\eta_t\sqrt{\omega+\sum_{i=1}^{q}\alpha_{i}y_{t-i}^2},
\quad t=0, \pm1, ...,
\end{flalign}
where all notations are inherited from model (\ref{dar_model}) except that $\alpha_{i}\geq 0$, and the conditional mean and volatility specifications can have different orders $p$ and $q$.
With these exceptions,
we are able to cope with the over-parameterization problem by checking whether
some coefficients are significant from zero in model (\ref{adar_model}).
 However, this makes the statistical inference of model (\ref{adar_model}) non-standard, since the volatility coefficient $\alpha_{i}$ is allowed to lie on the boundary of the parameter space (see, e.g.,
 \cite{gourieroux1982likelihood}, \cite{andrews1999estimation, andrews2001testing},
\cite{francq2007quasi, francq2009testing},
 \cite{iglesiaslinton2007}, \cite{cavaliere2017}
 and \cite{pedersen2017inference}). Also, when $\alpha_{i}$ is allowed to be zero,
 \cite{francq2007quasi} has demonstrated that
the GQMLE of the ARCH model (i.e., model (\ref{adar_model}) with $u=0$ and $\phi_{i}\equiv 0$)
requires a finite sixth moment of $y_{t}$ for its asymptotics, and
this makes the GQMLE of model (\ref{adar_model}) deficient to handle the heavy-tailed data
with an infinite sixth moment in many circumstances.

%

This paper contributes to the literature in three aspects. First,
a self-weighted GQMLE (S-GQMLE) is proposed for model (\ref{adar_model}) and its limiting distribution is shown to be a projection of a normal vector onto a convex cone by a quadratic approximation.
Based on this S-GQMLE, the Wald, Lagrange multiplier and quasi-likelihood ratio
tests are constructed to examine the nullity of some coefficients; their limiting distributions are established
under both null and local alternative hypotheses,  and their power performance is investigated under  local alternative hypotheses.
As a special
interest, testing for the null hypothesis of one coefficient equaling to zero is also studied.
By allowing for the null volatility coefficients,
the estimation and testing based on the GQMLE for the conditional
variance models have been well known for their non-standard asymptotics (see, e.g., \cite{andrews1999estimation, andrews2001testing}, \cite{francq2007quasi, francq2009testing} and \cite{pedersen2017inference}), but fewer attempts have been made to study their
asymptotics in the presence of the conditional mean structure. Our study on the S-GQMLE and its related tests for model (\ref{adar_model})
fills this gap. Interestingly, we find that even when the null volatility coefficients exist,
the S-GQMLE of the conditional mean parameter in model (\ref{adar_model}) is always asymptotically normal, and this property generally does not hold for the ARMA-GARCH model.
Hence, if we only examine the nullity of the conditional mean coefficients in model (\ref{adar_model}), the Wald, Lagrange multiplier and quasi-likelihood ratio tests can be implemented with standard asymptotics. In contrast, when
the volatility coefficients are included for nullity examination, the asymptotics of these three tests become non-standard. In view of
this important feature of model (\ref{adar_model}), we can use these three  tests to first
detect the nullity of the conditional mean coefficients by standard asymptotics, and then detect the nullity of the volatility coefficients by non-standard asymptotics. We shall emphasize that the preceding two-step procedure is not applicable
for the ARMA-GARCH model in general, since the distribution of the GQMLE of their conditional mean parameter is indeed non-standard caused by the null volatility coefficients.

Second, motivated by \cite{wongling2005}, we propose a new mixed portmanteau test to check the adequacy of model (\ref{adar_model}).
Diagnostic checking for model adequacy is important in time series analysis.
The seminal work in \cite{ljungbox1978} constructed a portmanteau test for the conditional mean model, and
later a similar portmanteau test was developed for the volatility model in \cite{limak1994}. Both portmanteau tests and their
many variants have the standard chi-squared limiting null distribution; see, e.g., \cite{zhu2016} and references therein.
When the null volatility coefficients are allowed in model (\ref{adar_model}), we find that
our mixed portmanteau test has a non-standard limiting null distribution, which is not
the standard chi-squared distribution any more. This result is new to the literature, and it
reveals that the null volatility coefficients have a non-ignorable effect on the model diagnostic checking.
To implement our mixed portmanteau test in practice,
we shall apply the Wald, Lagrange multiplier and quasi-likelihood ratio tests to obtain a reduced ADAR model with all positive volatility coefficients, and then use the standard chi-squared limiting null distribution for our mixed test.


Third, our entire statistical inference procedure aforementioned is valid as long as $y_{t}$ is stationary, and hence it can have a wide applicable scope in dealing with the heavy-tailed data.
Heavy-tailedness is often observed in many empirical data (see, e.g., 
\cite{rachev2003}, \cite{hill2015} and \cite{zhu2015lade}). When the null volatility coefficients exist in ARCH-type models,
the statistical inference methods in \cite{francq2007quasi, francq2009testing}
and \cite{pedersen2017inference} require
$y_t$ to have a finite sixth moment.
In contrast, our entire methodologies have no moment restriction on $y_{t}$ resulting from the use of the S-GQMLE, which is motivated by the self-weighting technique in \cite{ling2005self}. The self-weighting technique is necessary only when $y_t$ has an infinite sixth moment, and its idea is to apply the self-weight functions to reduce the effect of leverage data so that no moment condition of $y_t$ is needed. We emphasize that the ARMA-GARCH model with the S-GQMLE in
\cite{ling2007self} is also applicable to the heavy-tailed data. However, the asymptotics of the S-GQMLE in
\cite{ling2007self} do not allow null volatility coefficients, and hence no statistical inference method is proposed
in the presence of the null volatility coefficients. Finally,
the importance of our entire methodologies is illustrated by simulation studies and one real example.


The remainder of the paper is organized as follows.
Section \ref{swqmle} presents the S-GQMLE and establishes its asymptotics.
Section \ref{testing} constructs three tests to test for the null coefficients and obtains
their asymptotics.
Section \ref{poweranalysis} analyzes the power of these three tests.
Section \ref{modelchecking} proposes a portmanteau test for the model diagnostic checking.
Simulation results are reported in Section \ref{simulation}, and one real example is given in
Section \ref{emexample}.
Technical proofs of all theorems are relegated to Appendices.

Throughout the paper, $A'$ is the transpose of a matrix $A$,
 $\|A\|=(\mathrm{tr}(A'A))^{1/2}$ is the Frobenius norm of a matrix $A$,
$\langle x,y\rangle_A=x'Ay$ for any $x,y\in\mathbb{R}^{s}$ is the inner product induced by a positive
definite matrix $A\in\mathbb{R}^{s\times s}$, $\|x\|^2_{A}=\langle x,x\rangle_A$ is the norm of $x\in
\mathbb{R}^{s}$, $\Phi(\cdot)$ is the c.d.f. of standard normal random variable,
$I(\cdot)$ is the indicator function, and
$\rightarrow_\mathcal{{L}}$ denotes the convergence in distribution.

\section{Self-weighted Gaussian quasi-maximum likelihood estimation}\label{swqmle}

Let $\theta=(\phi', \alpha')'\in\mathbb{R}^{d}$ be the unknown parameter of model (\ref{adar_model}), where
$\phi=(u, \phi_1,...,\phi_p)'$, $\alpha=(\omega, \alpha_1,...,\alpha_q)'$ and $d=p+q+2$.  Let $m=\max(p,q)$.
Assume that the observations $\{y_{-m}, ..., y_n\}$ are generated from model (\ref{adar_model})
with the true value
$\theta_0=(\phi_0', \alpha_0')'$, where $\phi_0=(u_{0}, \phi_{10},...,\phi_{p0})'$ and
$\alpha_0=(\omega_0, \alpha_{10},...,\alpha_{q0})'$. Given the observations $\{y_{-m}, ..., y_n\}$, the self-weighted Gaussian quasi-maximum likelihood estimator (S-GQMLE) of $\theta_{0}$ is ${\hat\theta}_n:=(\hat{\phi}_n',\hat{\alpha}_{n}')'$, which is
defined as
\begin{flalign}\label{qmle}
{\hat\theta}_n=\arg\min_{\theta\in\Theta}F_n(\theta)
:=\arg\min_{\theta\in\Theta}\frac{1}{n}\sum_{t=1}^nw_t\ell_t(\theta),
\end{flalign}
where $\Theta$ is the parameter space,
$w_t:= w(y_{t-1},...,y_{t-m})$ is the self-weighted function with
$w(\cdot)$ being a measurable real positive and bounded function on
$\mathbb{R}^{m}$, and
\begin{equation}\label{likelihood}
\ell_t(\theta)=\frac{1}{2}\left\{\ln(\alpha'\mathbf{x}_{t-1})
+\frac{(y_t-\phi'\mathbf{y}_{t-1})^2}{\alpha'\mathbf{x}_{t-1}}\right\}
\end{equation}
with $\mathbf{y}_t=(1, y_{t},...,y_{t-p+1})'$ and $\mathbf{x}_t=(1, y_{t}^2,..., y_{t-q+1}^2)'$. Particularly, when $w_t=1$, the S-GQMLE reduces to the classical GQMLE in \cite{ling2004estimation}.

To obtain the asymptotic properties of ${\hat\theta}_n$, the following four assumptions are needed.

\begin{ass}\label{a.1}
$\{y_t\}$ is strictly stationary and ergodic.
\end{ass}

\begin{ass}\label{a.2}
The parameter space $\Theta$ is compact with $|u|<\bar{u}$, $|\phi_i|<\bar{\phi}$, $i=1,2,\cdots,p$,
$\underline{\omega}\leq\omega\leq \bar{\omega}$, and $0\leq\alpha_j\leq\bar{\alpha}$, $j=1,2,\cdots,q$, where $\bar{u}$, $\bar{\phi}$, $\underline{\omega}$, $\bar{\omega}$, and $ \bar{\alpha}$ are all finite positive constants.
\end{ass}

\begin{ass}\label{a.3}
 $E\{(w_t+w_t^2)(\left\|\mathbf{z}_{t-1}\right\|^2+\left\|\mathbf{z}_{t-1}\right\|^3)\}<\infty$,
  where $\mathbf{z}_{t-1}=(1,y_{t-1}^2,\cdots,y_{t-m}^2)'$.
\end{ass}

\begin{ass}\label{a.4}
The matrix $D=\left(\begin{matrix}
1& \tfrac{\kappa_3}{\sqrt{2}}\\
\tfrac{\kappa_3}{\sqrt{2}} &\tfrac{\kappa_{4}-1}{2}
\end{matrix}
\right)$ is positive definite, where $\kappa_3=E\eta_t^{3}$ and $\kappa_4=E\eta_t^4<\infty$.
\end{ass}

We offer some remarks on the aforementioned assumptions.
Assumption \ref{a.1} is a mild setting for time series models.
When $p=q=1$, a sufficient and necessary condition for Assumption \ref{a.1} was obtained in
\cite{bork2001Kl} and \cite{chenliling2014}. When $p=q>1$, a sufficient yet complicated condition for Assumption \ref{a.1} is available in  \cite{ling2007double}.

\begin{figure}[!htbp]
\centering
\includegraphics[scale=0.5]{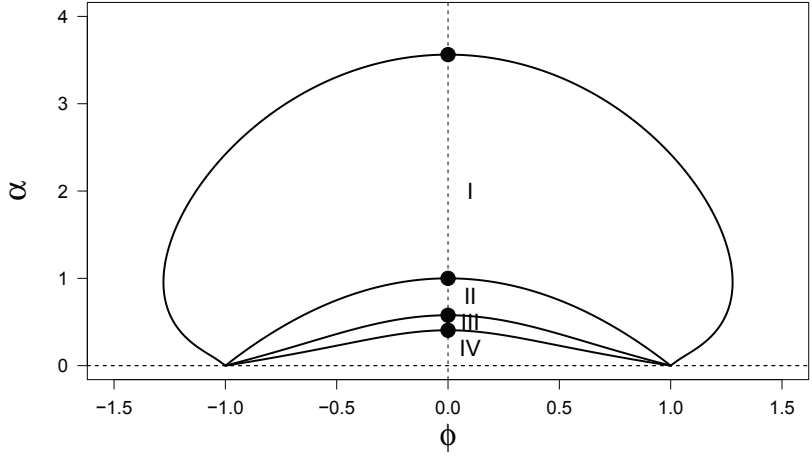}
\caption{Regions of strict stationarity (I+II+III+IV), 2nd moment (II+III+IV), 4th moment (III+IV) and
6th moment (IV) of $y_t$ in model (\ref{darone}).}\label{stationary}
\end{figure}

Assumption \ref{a.2} allows the volatility coefficient $\alpha_i$ to be zero.
In \cite{ling2004estimation, ling2007double},
each $\alpha_{i}$ is required to have a positive lower bound so that the
GQMLE only needs a finite fractional moment of $y_{t}$ for its asymptotic normality.
However, the requirement that $\alpha_{i}>0$ for each $i$ is stringent
and could cause the trouble of over-parameterization.
Under Assumption \ref{a.2}, the over-parameterization problem can be solved, but
as a trade-off, the asymptotic distribution of the GQMLE becomes non-standard
and requires $Ey_t^6<\infty$  (see, e.g.,
\cite{francq2007quasi} and \cite{pedersen2017inference}). In applications,
the finiteness of $Ey_t^6$ could be restrictive for two reasons.
First, this moment condition does not allow us to deal with many heavy-tailed data. Second,
this moment condition gives us a small admissible parameter space.
As a simple illustration, we consider a DAR($1, 1$) model:
\begin{flalign}\label{darone}
y_t=\phi y_{t-1}+\eta_t\sqrt{\omega+\alpha y_{t-1}^2}.
\end{flalign}
When $\eta_t\sim\mathcal{N}(0, 1)$, Table \ref{constraints} gives the constraints on the parameter $(\phi, \alpha)$ for strict stationarity,
the 2nd, 4th, and 6th moments of $y_t$,
and Fig\,\ref{stationary} displays these constraints graphically.
From this figure, we can see that the region of 6th moment is much smaller than
that of strict stationarity.
Hence, it is practically important to release the moment condition of $y_{t}$ so that the admissible parameter space is enlarged as much as possible.

\begin{table}[!htbp]
	\centering
	\caption{Parameter constraints for model $(\ref{darone})$}
	\label{constraints}
	\begin{tabular}[here]{lcccr}
		\firsthline
		\multicolumn{1}{l}{Condition}& Constraint&Region in Fig\,\ref{stationary}\\
		\hline
 Strict stationarity& $E\log|\phi+\eta_t\sqrt{\alpha}|<0$& I+II+III+IV\\
 \hline
 $Ey_t^2<\infty$ &$\phi^2+\alpha<1$&II+III+IV\\
 \hline
 $Ey_t^4<\infty$ &$\phi^4+6\phi^2\alpha+3\alpha^2<1$&III+IV\\
 \hline
 $Ey_t^6<\infty$&$ \phi^6+15\phi^4\alpha+45\phi^2\alpha^2+15\alpha^3<1$&IV\\
		\hline
	\end{tabular}
\end{table}

Assumption \ref{a.3} plays a key role in releasing the moment condition of $y_t$.
When $Ey_{t}^{6}<\infty$, it is valid without the weight (i.e., $w_{t}=1$).
When
$Ey_{t}^{6}=\infty$, the weight $w_t$ is introduced to reduce the effect of leverage points by shrinking their weights
on the objective function $F_n(\theta)$ so that no moment condition of $y_t$
is needed but at the sacrifice of  efficiency.
This idea was initiated by \cite{ling2005self}, and it has been adopted in many studies; see, e.g.,
\cite{ling2005self,ling2007self}, \cite{pan2007}, \cite{fz2010}, \cite{zhu2011global,zhu2015lade} and \cite{yang2017self}.
In practice, the selection of $w_t$ is similar to that of the influence function in
\cite{huber1996robust}.
For example, we can follow \cite{horvath2004} to choose
\begin{eqnarray}\label{weight3}
w_t=\frac{1}{1+\sum_{i=1}^{m}y_{t-i}^6}
\end{eqnarray}
or we can follow \cite{ling2005self} to choose
\begin{eqnarray}\label{weight2}
w_t=
\begin{cases}
1, & \mbox{if $a_t=0$, } \\
C_w^3/a_t^3, &\mbox{if $a_t\neq 0$, }
\end{cases}
\end{eqnarray}
where $a_t=\sum_{i=1}^{m}y_{t-i}^2I(y_{t-i}^2\geq C_w)$ for some constant $C_w>0$, and $C_w$ is chosen as the 90\%  or 95\% percentile of $\{y_t^2\}_{t=1}^{n}$, empirically.
However, when the second moment of $y_t$ does not exist,  the 95\% empirical percentile of $\{y_t^2\}_{t=1}^{n}$ might be very large, leading to malfunction of (\ref{weight2}). Therefore, we prefer to use $w_t$ in (\ref{weight3}) subsequently, but
leave the selection of the optimal $w_t$ as an open problem.

Assumption \ref{a.4} is general to derive the asymptotic distribution of $\hat{\theta}_n$. As shown in \cite{wilkins1944note}, this assumption is equivalent to that
$\mathbb{P}(\eta_t^2-c\eta_t=1)<1$ for any $c\in\mathbb{R}$, which is satisfied for continuous $\eta_{t}$.

Next, let $J=E(w_t\Gamma_t(\theta_0)\Gamma_t(\theta_0)')$ and $\Sigma=E(w_t^2\Gamma_t(\theta_0)D\Gamma_t(\theta_0)')$ with
\begin{eqnarray*}
 \Gamma_t(\theta)=\left(\begin{matrix}
		\frac{\mathbf{y}_{t-1}}{\sqrt{(\alpha'\mathbf{x}_{t-1})}} & 0_{(p+1)\times 1}\\	
		0_{(q+1)\times 1}& \frac{\mathbf{x}_{t-1}}{\sqrt{2}(\alpha'\mathbf{x}_{t-1})}
			\end{matrix}\right).
\end{eqnarray*}
We are ready to give our first main result on the consistency and asymptotic distribution of $\hat{\theta}_n$.

\begin{thm}\label{mainthm}
Suppose that Assumptions \ref{a.1}-\ref{a.3} hold. Then,

$\mathrm{(i)}$ $\hat{\theta}_n\rightarrow\theta_0$ almost surely (a.s.) as $n\to\infty$;

\noindent

$\mathrm{(ii)}$ if Assumption \ref{a.4} further holds, $\sqrt{n}(\hat{\theta}_n-\theta_0)\rightarrow_{\mathcal{L}}\lambda^{\Lambda}:=\arg \inf_{\lambda\in\Lambda}
\|Z-\lambda\|_{J}$
as $n\rightarrow \infty$,

\noindent where
$Z\sim \mathcal{N}(0,J^{-1}{\Sigma}J^{-1})$ and $\Lambda:=\Lambda_1\times\Lambda_2\times\cdots\times\Lambda_{p+q+2}$
with $\Lambda_i=\mathbb{R}$ for $ i=1,2,\cdots,p+2$, and
\begin{flalign*}
\Lambda_{p+j+2}=
\left\{
\begin{array}{ll}
[0,\infty), &  \mbox{ if }\alpha_{j0}=0, \\
\mathbb{R},  &  \mbox{ if }\alpha_{j0}\neq0,
\end{array}
\right. \mbox{ for }j=1,2,\cdots,q.
\end{flalign*}
\end{thm}

Theorem \ref{mainthm} implies that when $\theta_0$ is not an interior point of $\Theta$ (i.e., some of its volatility coefficients are on the boundary),
the limiting distribution of $\hat{\theta}_{n}$ is no longer Gaussian but a projection of
a Gaussian random variable $Z$ onto the convex cone $\Lambda$	
with a metric induced  by the inner product $\langle\cdot,\cdot\rangle_J$.
The uniqueness of such a projection is guaranteed by the convexity of $\Lambda$.
Particularly, when $\theta_0$ is an interior point of $\Theta$, we have $\Lambda=\mathbb{R}^{p+q+2}$ and then $\lambda^{\Lambda}=Z\sim \mathcal{N}(0,J^{-1}\Sigma J^{-1})$.

Write $Z=(Z_{\phi}',Z_{\alpha}')'$ and
	$\lambda^\Lambda=(\lambda_{\phi}^{\Lambda'},\lambda_{\alpha}^{\Lambda'})'$.
In view of that $J=\mathrm{diag}\{J_{\phi},J_{\alpha}\}$ is a block diagonal matrix, by Theorem 4 in \cite{andrews1999estimation}, it is not hard to see that
\begin{flalign}\label{2_6}
\lambda_{\phi}^{\Lambda}=\inf_{\lambda_{\phi}\in\Lambda_{\phi}}
	\|Z_{\phi}-\lambda_{\phi}\|_{J_{\phi}}=Z_{\phi}\,\,\,\mbox{ and }\,\,\,
\lambda_{\alpha}^{\Lambda}=\inf_{\lambda_{\alpha}\in\Lambda_{\alpha}}\|Z_{\alpha}-\lambda_{\alpha}\|_{J_{\alpha}},
\end{flalign}
where
	$J_{\phi}=E\{w_t\mathbf{y}_{t-1}\mathbf{y}_{t-1}'/(\alpha_0'\mathbf{x}_{t-1})\}$, $J_{\alpha}=E\{w_t\mathbf{x}_{t-1}\mathbf{x}_{t-1}'/(\alpha_0'\mathbf{x}_{t-1})^2\}/2$, $\Lambda_{\phi}=\mathbb{R}^{p+1}$, and $\Lambda_{\alpha}=\Lambda_{p+2}\times\cdots\times\Lambda_{p+q+2}$.
The result (\ref{2_6}) implies that $\hat{\phi}_{n}$ is always asymptotically normal,
no matter whether the null volatility coefficients exist. This important feature guarantees that we can examine the significance of the conditional mean coefficients by using the standard statistical inference methods, and then implement the statistical inference for the significance of the volatility coefficients. The validity of this two-step procedure is mainly because the matrix $J$ is block diagonal, and
it does not need $\Sigma$ to be block diagonal, allowing $\hat{\phi}_n$ and $\hat{\alpha}_n$ to be asymptotically correlated.
Note that the matrix $J$ is the expectation of the Hessian matrix of the objective function. For the ARMA-GARCH model, the corresponding matrix is not diagonal in general, and hence the
GQMLE of the conditional mean parameter may not be asymptotically normal if the null volatility coefficients exist.

\section{Testing for null coefficients}\label{testing}
In this section, we consider the Wald, Lagrange multiplier (LM) and
quasi-likelihood ratio (QLR) tests to detect whether some coefficients  are equal to zero in model (\ref{adar_model}).
Since the significance of the conditional mean coefficients can be examined ahead, we only focus on the
tests for the null volatility coefficients.


We split the true parameter $\theta_0$ into three parts such that
$\theta_0=(\theta_0^{(1)'},\theta_0^{(2)'},\theta_0^{(3)'})'$,
where $\theta_0^{(i)}\in \mathbb{R}^{d_{i}}$ for $i=1, 2, 3$, and $d=d_{1}+d_{2}+d_{3}$.
Without loss of generality, we assume $\theta_0^{(1)}=\phi_0\cup\{\alpha_{0i}:\alpha_{0i}>0\}$, $\theta_0^{(2)}=0_{d_2\times 1}$ and $\theta_0^{(3)}=0_{d_3\times1}$.
 That is, $\theta_0^{(1)}$ contains the conditional mean coefficients as well as  volatility coefficients that are  strictly larger than zero,
 and $(\theta_0^{(2)'},\theta_0^{(3)'})'$ contains all volatility coefficients on the boundary. Note that from now on, the order of components in $\mathbf{x}_t$ and $\mathbf{y}_t$
is changed according to the  splitting way of $\theta_0$.
Let  $K=(0_{(d-d_1)\times d_1},I_{d-d_1})$ and $K_{\alpha}=(0_{d_3\times(d-d_3)}, I_{d_3})$.
Our null hypothesis is set as
\begin{flalign*}
H_0:\theta_{0}^{(3)}=0_{d_3\times1}\,\,\,(\mbox{i.e.}, \,K_{\alpha}\theta_0=0_{d_3\times1}).
\end{flalign*}
Under $H_0$, the nuisance coefficients vector $(\theta_0^{(1)'},\theta_0^{(2)'})'$
allows $\theta_0^{(2)}$ on the boundary. This setting is similar to that in \cite{pedersen2017inference}, and
more general than that in \cite{francq2009testing}, which only considers the case of $d_{2}=0$.


To construct our test statistics, the following notations are needed:
\begin{flalign*}
{J}_{n}(\theta)&=\frac{\partial^2 F_n({\theta})}{\partial\theta\partial\theta '},\quad {\Sigma}_{n}(\theta)=\frac{1}{n}\sum_{t=1}^{n}w_t^2\Gamma_t(\theta)'{D}_{n}(\theta)\Gamma_t(\theta),\quad\mbox{and}\quad\\
{D}_{n}(\theta)&=\left(
\begin{matrix}
1 &
\frac{1}{\sqrt{2}\bar{w}n}\sum\limits_{t=1}^{n}\frac{w_t\epsilon_t^3(\phi)}{(\alpha'\mathbf{x}_t)^{3/2}}\\ \frac{1}{\sqrt{2}\bar{w}n}\sum\limits_{t=1}^{n}\frac{w_t\epsilon_t^3(\phi)}{(\alpha'\mathbf{x}_t)^{3/2}} &\frac{1}{2\bar{w}n}\sum\limits_{t=1}^{n}\frac{w_t\epsilon_t^4(\phi)}{(\alpha'\mathbf{x}_t)^{2}}-\frac{1}{2}
\end{matrix}
\right),
\end{flalign*}
where $\epsilon_t(\phi)=y_t-\phi'\mathbf{y}_{t-1}$ and $\bar{w}=n^{-1}\sum_{t=1}^{n}w_t$. With these notations, we denote
\begin{flalign*}
\hat{J}_n&=J_{n}(\hat{\theta}_n),\qquad \hat{\Sigma}_{n}=\Sigma_{n}(\hat{\theta}_n),\qquad \hat{D}_n=D_{n}(\hat{\theta}_n),\\
\hat{J}_{n|3}&=J_{n}(\hat{\theta}_{n|3}),\quad \hat{\Sigma}_{n|3}=\Sigma_{n}(\hat{\theta}_{n|3}),\quad
\hat{D}_{n|3}=D_{n}(\hat{\theta}_{n|3}),
\end{flalign*}
where $\hat{\theta}_{n|3}$ is the restricted S-GQMLE under $H_0$.
%
Our Wald, LM and QLR test statistics are defined as
\begin{flalign*}
W_n&=n\hat{\theta}_n^{(3)'}\{K_{\alpha}\hat{J}_{n}^{-1}\hat{\Sigma}_n\hat{J}_{n}^{-1}K_{\alpha}'\}^{-1}\hat{\theta}_n^{(3)},\\
L_n&=n\frac{\partial F_n(\hat{\theta}_{n|3})}{\partial \theta'}\hat{J}_{n|3}^{-1}K_{\alpha}'\{K_{\alpha}\hat{J}_{n|3}^{-1}\hat{\Sigma}_{n|3}\hat{J}_{n|3}^{-1}K_{\alpha}'\}^{-1}
K_{\alpha}\hat{J}_{n|3}^{-1}\frac{\partial F_n(\hat{\theta}_{n|3})}{\partial\theta},\\
Q_n&=2n\big[F_n(\hat{\theta}_{n|3})-F_n(\hat{\theta}_n)\big],
\end{flalign*}
respectively, and their limiting null distributions are given in the following theorem.

\begin{thm}\label{testmain}
Suppose that Assumptions \ref{a.1}-\ref{a.4} hold. Then, under $H_0$, as $n\to\infty$,

$\mathrm{(i)}$ $W_n\rightarrow_{\mathcal{L}} W:=\lambda^{\Lambda'}\Omega\lambda^{\Lambda}$;

$\mathrm{(ii)}$ $L_n\rightarrow_{\mathcal{L}} L:=\chi_{d_3}^2$;

$\mathrm{(iii)}$ $Q_n\rightarrow_{\mathcal{L}} Q:= \|Z-\lambda_{|3}^{\Lambda}\|^2_{J}-\|Z-\lambda^{\Lambda}\|^2_{J}=\lambda^{\Lambda'}\Xi\lambda^{\Lambda}-\lambda_{|3}^{\Lambda'}\Xi\lambda_{|3}^{\Lambda'}$,

\noindent where $\lambda^{\Lambda}$ and $Z$ are defined as in Theorem \ref{mainthm},
\begin{equation}\label{omega}
\Omega=K_{\alpha}'\{K_{\alpha} J^{-1}\Sigma J^{-1}K_{\alpha}'\}^{-1}K_{\alpha},\quad
\Xi=K'(KJ^{-1}K')^{-1}K,\quad\mbox{and}\quad
\lambda_{|3}^{\Lambda}=\arg\inf_{\lambda_{|3}\in\Lambda_{|3}}\|Z-\lambda_{|3}\|^2_{J}
\end{equation}
with $\Lambda_{|3}:=\mathbb{R}^{d_1}\times[0,\infty)^{d_2}\times\{0\}^{d_3}$.
Particularly, when $d_2=0$,  $Q=\lambda^{\Lambda'}\Xi\lambda^{\Lambda}$.
\end{thm}



Theorem \ref{testmain} shows that except for $L_n$, the limiting null distributions of $W_n$ and $Q_n$ are not intuitive.
The test $L_n$ has standard chi-squared limiting null distribution, since its limiting distribution depends on that of
the score $\frac{\partial F_n(\theta_{0})}{\partial \theta}$, which is asymptotically normal under $H_0$ (see (\ref{gamma}) in Appendix A). On the contrary,
the tests $W_n$ and $Q_n$ have non-standard limiting null distributions, since their limiting distributions rely on that of
$\hat{\theta}_n^{(3)}$, however, $\hat{\theta}_n^{(3)}$ is not asymptotically normal under $H_0$ as shown in Theorem \ref{mainthm}(ii).

By letting
\begin{equation}\label{est_omega_xi}
\hat{\Omega}_n=K_{\alpha}'\{K_{\alpha} \hat{J}_n^{-1}\hat{\Sigma}_n \hat{J}_n^{-1}K_{\alpha}'\}^{-1}K_{\alpha}\quad\mbox{and}\quad
\hat{\Xi}_n=K'(K\hat{J}_n^{-1}K')^{-1}K
\end{equation}
be the estimators of  $\Omega$ and $\Xi$ in (\ref{omega}),
we propose an algorithm, which is similar to Algorithm 1 in \cite{pedersen2017inference}, to calculate the critical values
of $W_n$ and $Q_n$ in practice.
\begin{alg}\label{alg1}
 {\bf $($Simulated critical values of $W_n$ and $Q_n$$)$}
\begin{enumerate}
\item Draw $\epsilon_*$ from $\mathcal{N}(0, I_{d})$ and then compute $Z_*=[\hat{J}_n^{-1}\hat{\Sigma}_n\hat{J}_n^{-1}]^{1/2}\epsilon_*$.
\item Find $\lambda_*^{\Lambda}$ that minimizes $(Z_*-\lambda_*)'\hat{J}_n(Z_*-\lambda_*)$ for $\lambda_* \in \Lambda$ and
 $\lambda_{*|3}^{\Lambda}$ that minimizes $(Z_*-\lambda_{*|3})'\hat{J}_n(Z_*-\lambda_{*|3})$ for $\lambda_{*|3} \in \Lambda_{|3}$.
\item Calculate $w_*=\lambda_*^{\Lambda'}\hat{\Omega}_n\lambda_*^{\Lambda}$ and $q_*=\lambda_*^{\Lambda'}\hat{\Xi}_n\lambda_*^{\Lambda}-\lambda_{*|3}^{\Lambda'}\hat{\Xi}_n\lambda_{*|3}^{\Lambda'}$.
\item Repeat steps 1-3 $N$ times to get $\{w_{i,*}\}_{i=1}^{N}$ and $\{q_{i,*}\}_{i=1}^{N}$, where
$w_{i,*}$ and $q_{i,*}$ are the realizations of $w_*$ and $q_*$ in $i^{th}$ time, respectively.
At the level $\beta\in(0, 1)$, the  critical values of $W_{n}$ and $Q_{n}$ are
the empirical $100(1-\beta)\%$ sample percentiles of
$w_*$ and $q_*$
based on $\{w_{i,*}\}_{i=1}^{N}$ and $\{q_{i,*}\}_{i=1}^{N}$, respectively.
\end{enumerate}
\end{alg}

To implement Algorithm \ref{alg1}, we need know $\Lambda$ and $\Lambda_{|3}$. Hence, Algorithm \ref{alg1} is only
applicable when either $d_2=0$ (i.e., no nuisance parameters on the boundary exist under $H_0$) or $d_2\not=0$ with a known
$\theta_0^{(2)}$ (i.e., the true nuisance parameters on the boundary are known under $H_0$).
If $d_2\not=0$ with an unknown
$\theta_0^{(2)}$, so far it is unclear how to obtain the critical values of $W_n$ and $Q_n$. In practice,
we recommend to use both $W_n$ and $Q_n$ by presuming $d_2=0$. If $H_0$ is rejected in this case, we then get a reduced ADAR model
with coefficients $(\theta_0^{(1)'},\theta_0^{(2)'})'$, from which we could get the supportive evidence of
$d_2=0$ if
our tests imply that all coefficients
 $(\theta_0^{(1)'},\theta_0^{(2)'})'$ are non-zeros.

Besides the simulation method in  Algorithm \ref{alg1}, one may  use the ``$m$ out of $n$ bootstrap'' method
as in \cite{politis1994} and \cite{andrews2000} to compute the critical values of $W_n$ and $Q_n$. Although this bootstrap method is valid in theory,
its performance could be sensitive to the choice of the subsampling size $m$ (as demonstrated by our un-reported simulation results),
while  how to choose the optimal $m$ remains unsolved in our time series setting. Based on this argument, we recommend to use
Algorithm \ref{alg1} for convenience.

In many situations, we are mostly interested in testing the nullity of one volatility coefficient:
\begin{equation*} 
H_0': \alpha_{i0}=0\quad \mbox{for some given $i\in\{1,2,\cdots,q\}$}.
\end{equation*}
Besides the Wald, LM and QLR tests, a $t$-type test statistic $t_n$ is often used in practice to detect
$H_{0}'$, where
\begin{flalign*}
t_n=\frac{\hat{\alpha}_{in}}{\hat{\sigma}_{\hat{\alpha}_{in}}},
\end{flalign*}
and $\hat{\sigma}_{\hat{\alpha}_{in}}$ is the square root of the $i$th diagonal element of $\hat{J}_n^{-1}\hat{\Sigma}_n\hat{J}_n^{-1}$. Particularly, when $d_3=1$, we have $t_n^2=W_n$.

When $d_2>0$ with a known $\theta_0^{(2)}$, we can  apply
Algorithm \ref{alg1} to find the critical values of $t_{n}$, $W_{n}$ and $Q_{n}$.
When $d_2=0$, $t_{n}$, $W_{n}$ and $Q_{n}$ have
simpler critical regions, which are closely related to standard normal and chi-squared distributions.

\begin{thm}\label{one coefficient}
Suppose that Assumptions \ref{a.1}-\ref{a.4} hold and $d_2=0$. Then,
  the $t$-type, Wald, LM and QLR tests of asymptotic significance level $\beta\in (0, 1/2)$ for $H_0'$ are defined by the critical regions
\begin{flalign*}
\{t_n>\Phi^{-1}(1-\beta)\},\qquad \{W_n>\chi_{1,1-2\beta}^2\}, \qquad
 \{L_n>\chi^2_{1,1-\beta}\}, ~\qquad  \{\hat{\xi}_nQ_n>\chi^2_{1,1-2\beta}\},
\end{flalign*}
respectively, where $\hat{\xi}_n=\hat{\Omega}_n/\hat{\Xi}_n$, and $\hat{\Omega}_n$ and $\hat{\Xi}_n$ are defined in (\ref{est_omega_xi}).
\end{thm}

The preceding theorem demonstrates that  when the coefficient $\alpha_{i0}$ lies on the boundary of parameter space, the standard critical regions of
$t_{n}$, $W_{n}$ and $Q_{n}$ are not correct, except $L_n$. In other words, if we follow $L_n$ to use the conventional critical regions for
$t_{n}$, $W_{n}$ and $Q_{n}$, we would encounter a distorted size problem for the latter three tests.

\section{Power analysis}\label{poweranalysis}

This section studies the efficiency of the Wald, LM and QLR tests via the Pitman analysis.

First, we need the asymptotic distribution of the S-GQMLE
under sequences of local alternatives to the true parameter $\theta_0$.
Let $\theta_n=\theta_0+h/\sqrt{n}$, where  $h=(h_1,\cdots,h_{p+q+2})'$ with $(h_1,\cdots,h_{p+1})'\in\mathbb{R}^{p+1}$ and $(h_{p+2},\cdots,h_{p+q+2})\in (0,\infty)^{q+1}$ such that
$\theta_n\in\Theta$ for sufficiently large $n$.
When $n$ is sufficiently large, we can define a strictly stationary solution to
\begin{flalign*}
y_{t,n}=\left(u_{0}+\frac{h_{1}}{\sqrt{n}}\right)+\sum_{i=1}^{p}\left(\phi_{0i}+\frac{h_{i+1}}{\sqrt{n}}\right)y_{t-i,n}
+\eta_t\sqrt{\left(\omega_0+\frac{h_{p+2}}{\sqrt{n}}\right)+\sum_{j=1}^{q}\left(\alpha_{0j}+\frac{h_{p+j+2}}{\sqrt{n}}\right)y_{t-j,n}^2},
\end{flalign*}
where $\eta_t$ is defined as in model (\ref{adar_model}).
Based on $\{y_{-m,n},\cdots,y_{n,n}\}$, the S-GQMLE is
\begin{equation}\label{local}
\hat{\theta}_{n,h}
=\arg\min_{\theta\in\Theta}\frac{1}{n}\sum_{t=1}^{n}w_{t,n}l_{t,n}(\theta),
\end{equation}
where
$$l_{t,n}(\theta)=\frac{1}{2}\left\{\ln(\alpha'\mathbf{x}_{t-1,n})
+\frac{\epsilon^2_{t,n}(\phi)}{\alpha'\mathbf{x}_{t-1,n}}\right\}$$
with $\epsilon_{t,n}(\phi)=y_{t,n}-\phi'\mathbf{y}_{t-1,n}$ and $\mathbf{y}_{t,n}=(1, y_{t,n},...,y_{t-p+1,n})'$.
Below, we impose a stronger sufficient assumption on the self-weighted function $w_{t,n}$ to derive the asymptotics of
$\hat{\theta}_{n,h}$.

\begin{thmbis}{a.3}\label{a.3p}
$(w_{t,n}+w_{t,n}^2)(\left\|\mathbf{z}_{t-1,n}\right\|^2+\left\|\mathbf{z}_{t-1,n}\right\|^{3+\delta})<\infty$ for some $\delta\geq 0$, where $\mathbf{z}_{t-1,n}=(1,y_{t-1,n}^2,\cdots,y_{t-m,n}^2)'$.
\end{thmbis}
\noindent

Denote by $\mathbb{P}_{n,h}$ the law of $y_{t,n}$. Then, we have the following result.
\begin{thm}\label{mainthm2}
Suppose that Assumptions \ref{a.1}-\ref{a.2} and \ref{a.4} hold. Then, under $\mathbb{P}_{n,h}$,

$\mathrm{(i)}$ if Assumption \ref{a.3p} holds with $\delta=0$, $\hat{\theta}_{n,h}\rightarrow\theta_0$ in probability as $n\rightarrow\infty$;

$\mathrm{(ii)}$ if Assumption \ref{a.3p} holds with $\delta=2$, $\sqrt{n}(\hat{\theta}_{n,h}-\theta_0)\rightarrow_\mathcal{{L}} \lambda^{\Lambda}(h)$ as $n\rightarrow\infty$,
where
\begin{flalign*}  
\lambda^{\Lambda}(h)=\arg\inf_{\lambda\in\Lambda}\|Z+h-\lambda\|_{J}
\end{flalign*} with
$Z$ being defined as in Theorem \ref{mainthm}.
\end{thm}
	
We emphasize that since the log-likelihood ratio is not asymptotically normal in our setting, it seems difficult to apply the classical
Le Cam's third lemma to prove Theorem  \ref{mainthm2}(ii); see \cite{francq2009testing} for more discussions. In this paper,
we show Theorem  \ref{mainthm2}(ii) in a direct way.


Let $\chi^2_s(c)$ be the noncentral chi-squared distribution with noncentrality parameter $c$ and
degrees of freedom $s$. The
asymptotic distributions of all three test statistics under the local
alternatives are given as follows.

\begin{thm}\label{locallimit}
Suppose that the conditions in Theorem \ref{mainthm2}(ii) hold and $\theta_{0}^{(3)}=0_{d_{3}\times 1}$. Then, under $\mathbb{P}_{n,h}$, as $n\to\infty$,

$\mathrm{(i)}$ $W_n\rightarrow_{\mathcal{L}}W(h):=\lambda^{\Lambda}(h)'\Omega\lambda^{\Lambda}(h)$;

$\mathrm{(ii)}$ $L_n\rightarrow_{\mathcal{L}}L(h):=\chi_{d_3}^2(h'\Omega h);$

$\mathrm{(iii)}$ $Q_n\rightarrow_{\mathcal{L}}Q(h):=\|Z+h-\lambda_{|3}^{\Lambda}\|^2_{J}-\|Z+h-\lambda^{\Lambda}\|^2_{J}
=\lambda^{\Lambda'}(h)\Xi\lambda^{\Lambda}(h)-\lambda_{|3}^{\Lambda'}(h)\Xi\lambda_{|3}^{\Lambda'}(h)$,

\noindent where $\Omega$ and $\Xi$ are defined in (\ref{omega}),  $\lambda^{\Lambda}(h)$ is defined as in Theorem \ref{mainthm2}, and
\begin{flalign*}
\lambda_{|3}^{\Lambda}(h)=\arg\inf_{\lambda\in\Lambda_{|3}}\|Z+h-\lambda\|_{J}
\end{flalign*}
with $Z$ being defined as in Theorem \ref{mainthm}.
\end{thm}

For all of four tests in Theorem \ref{one coefficient},
the following theorem shows that the local asymptotic power of the
$t$-type, Wald and QLR tests is the same, and it is higher than the one of LM test.

\begin{thm}\label{localpower}
Suppose that the conditions in Theorem \ref{mainthm2}(ii) hold and $d_2=0$. Then, the local asymptotic power of the $t$-type, Wald and QLR tests is
\begin{flalign*}  
\lim\limits_{n\rightarrow\infty} \mathbb{P}_{n,h}(t_n>\Phi^{-1}(1-\beta))=\lim\limits_{n\rightarrow\infty} \mathbb{P}_{n,h}(W_n>\chi^2_{1,1-2\beta}) =\lim\limits_{n\rightarrow\infty} \mathbb{P}_{n,h}(\hat{\xi}Q_n>\chi^2_{1,1-2\beta})=1-\Phi(c_1-h^*),
\end{flalign*}
and the local asymptotic power of LM test is
\begin{equation*}
\lim\limits_{n\rightarrow\infty}\mathbb{P}_{n,h}(L_n>\chi^2_{1,1-\alpha})=1-\Phi(c_2-h^*)+\Phi(-c_2-h^*)<1-\Phi(c_1-h^*),
\end{equation*}
where $h^*=h_d/\sigma_d$, $c_1=\Phi^{-1}(1-\beta)$ and $c_2=\Phi^{-1}(1-\beta/2)$.
\end{thm}

In addition to the Pitman analysis, the Bahadur slopes as in \cite{bahadur1960asymptotic} under fixed alternatives are also established for three test statistics in the supplementary material (\cite{jlz19}). However, as in \cite{francq2009testing}, a formal comparison of Bahadur slopes for all considered tests is not easy, since
$J$, $J_{0|3}$, $\Sigma$ and $\Sigma_{0|3}$ are unknown in closed form, particularly when the self-weighted function $w_{t}$ is included.


\section{Model checking}\label{modelchecking}
This section proposes a new portmanteau test to check the adequacy of model (\ref{adar_model}).
Define the self-weighted innovation $\zeta_t=w_t\eta_t$ and
 the self-weighted squared innovation $\xi_t=w_t^2(\eta_t^2-1)$.
 Accordingly, define the self-weighted residual $\hat{\zeta}_t=w_t\hat{\eta}_t$ and
 the self-weighted squared residual $\hat{\xi}_t=w_t^2(\hat{\eta}_t^2-1)$,
 where $\hat{\eta}_t=\epsilon_t(\hat{\phi}_n)/\sqrt{\hat{\alpha}_n'\mathbf{x}_{t-1}}$ is the residual of model (\ref{adar_model}).
 The idea of our mixed portmanteau test is
based on the fact that
 $\{\zeta_t\}$ (or $\{\xi_t\}$) is a sequence of uncorrelated random
variables. Hence, if model
 (\ref{adar_model}) is correctly specified,
it is expected that the sample autocorrelation function of  $\{\hat{\zeta}_t\}$ (or $\{\hat{\xi}_t\}$) at
lag $k$, denoted by $\hat{\rho}_{nk}$ (or $\hat{r}_{nk}$), is close to zero, where
\[
\hat{\rho}_{nk}=\frac{\sum_{t=k+1}^{n}(\hat{\zeta}_t-\bar{\zeta})
(\hat{\zeta}_{t-k}-\bar{\zeta})}{\sum_{t=1}^{n}(\hat{\zeta}_t-\bar{\zeta})^2}\,\,\,\mbox{ and }
\,\,\,
\hat{r}_{nk}=\frac{\sum_{t=k+1}^{n}(\hat{\xi}_t-\bar{\xi})
(\hat{\xi}_{t-k}-\bar{\xi})}{\sum_{t=1}^{n}(\hat{\xi}_t-\bar{\xi})^2},
\]
with $\bar{\zeta}=\sum_{t=1}^n\hat{\zeta}_t/n$ and $\bar{\xi}=\sum_{t=1}^n\hat{\xi}_t/n$.
On the other hand, if the value of $\hat{\rho}_{nk}$ (or $\hat{r}_{nk}$) deviates from zero significantly,
it implies that  the conditional mean (or variance) structure in model (\ref{adar_model}) is misspecified.
Motivated by this, our new mixed portmanteau test takes both $\hat{\rho}_{nk}$ and $\hat{r}_{nk}$ into account, and
this new test can detect the misspecification in conditional mean and variance simultaneously.

%
%
%
%
%

Let $M\geq1$ be a given integer, $\bar{\sigma}^2=(\kappa_4-1)Ew_t^4$,  $\hat{\rho}_{n}=(\hat{\rho}_{n1},\cdots,\hat{\rho}_{nM})'$, $\hat{r}_{n}=(\hat{r}_{n1},\cdots,\hat{r}_{nM})'$,
$U_{\rho}=(U_{\rho1}',\cdots,U_{\rho M}')'$, and $U_{r}=(U_{r1}',\cdots,U_{r M}')'$, where
\begin{flalign*}
U_{\rho k}=-\Big( E\Big[\frac{w_tw_{t-k}\eta_{t-k}\mathbf{y}'_{t-1}}{\sqrt{\alpha_0'\mathbf{x}_{t-1}}}\Big], 0_{1\times(q+1)}
\Big)\,\,\,\mbox{ and }\,\,\,
U_{rk}=-\Big(
0_{1\times(p+1)}, E\Big[\frac{w_t^2w_{t-k}^2(\eta_{t-k}^2-1)\mathbf{x}'_{t-1}}{\alpha_0'\mathbf{x}_{t-1}}\Big]
\Big).
\end{flalign*}
Meanwhile, let $\mathcal{G}$ be a mapping: $\mathbb{R}^{2M+p+q+2}\rightarrow\mathbb{R}^{2M}\times\Lambda$, defined as
\begin{flalign*}
\mathcal{G}(a)=\left(\begin{matrix}
a_1\\
a_2\\
\arg\inf\limits_{\lambda\in\Lambda}\|a_3-\lambda\|_{J}
\end{matrix}\right)
\end{flalign*}
for any $a=(a_1',a_2',a_3')'\in \mathbb{R}^{2M+d}$, where $a_1, a_2\in\mathbb{R}^{M}$ and $a_3\in\mathbb{R}^{d}$.
Then, we have the following result on the limiting joint  distribution of $(\hat{\rho}_n',\hat{r}_n')'$.

%
%
%

\begin{thm}\label{jointdistribution}
	Suppose that Assumptions \ref{a.1}-\ref{a.4} hold. Then, if model (\ref{adar_model}) is correctly specified,
as $n\to\infty$,
	\begin{equation}\label{joint_5_1}
	\sqrt{n}\left(
	\begin{matrix}
	\hat{\rho}_n\\
	\hat{r}_n
	\end{matrix}
	\right)
\rightarrow_{\mathcal{L}}V\mathcal{G}(g)
	\end{equation}
for some random vector $g\sim \mathcal{N}(0,G)$,
where \[
V=\left(
\begin{matrix}
I_M&0&(Ew_t^2)^{-1}U_\rho\\
0&I_M&(\bar{\sigma}^2)^{-1}U_r
\end{matrix}
\right)\,\,\,\mbox{ and }\,\,\,G=Ev_tv_t'
\]
with
$
v_t=\Big(\frac{\zeta_t\zeta_{t-1}}{Ew_t^2},\cdots,\frac{\zeta_t\zeta_{t-M}}{Ew_t^2},
        \frac{\xi_t\xi_{t-1}}{\bar{\sigma}^2},\cdots,\frac{\xi_t\xi_{t-M}}{\bar{\sigma}^2},J^{-1}w_t\frac{\partial\ell_t(\theta_0)}{\partial\theta}\Big).
$
\end{thm}

Based on Theorem \ref{jointdistribution}, we propose a new mixed portmanteau test statistic defined by
\begin{equation}\label{Q_m}
	Q_M=n\left(\begin{matrix}
	\hat{\rho}_n\\
	\hat{r}_n
	\end{matrix}\right)'(\hat{V}_n\hat{G}_n\hat{V}_{n}')^{-1}\left(\begin{matrix}
	\hat{\rho}_n\\
	\hat{r}_n
	\end{matrix}\right),
\end{equation}
where $\hat{V}_n$ and $\hat{G}_n$ are the consistent sample counterparts of $V$ and $G$, respectively.
By Theorem \ref{jointdistribution} and the continuous mapping theorem, we have
\begin{flalign*}
Q_M\rightarrow_{\mathcal{L}} \mathcal{G}(g)'V'(VGV')^{-1}V\mathcal{G}(g)\quad \mbox{as $n\to\infty$}.
\end{flalign*}
Hence, when some coefficients of $\theta_0$ are on the boundary of parameter space,
$\mathcal{G}(g)\not=g$ and our portmanteau test $Q_m$ in (\ref{Q_m}) has a non-standard limiting distribution;
when none of the coefficients of $\theta_0$ is on the boundary of parameter space,
$\mathcal{G}(g)=g$ and $Q_m$ has limiting distribution $\chi^2_{2M}$, which is
the standard result as many existing portmanteau tests. Note that
the non-standard limiting distribution of $Q_M$ depends on $\Lambda$, which is determined by the location of all
null volatility coefficients but is hard to be correctly specified. To implement $Q_M$ in practice, we need first
apply our Wald, LM and QLR tests to obtain a reduced ADAR model with all positive volatility coefficients, and
then use the standard chi-squared limiting null distribution for $Q_M$.
Even though this reduced ADAR model has no boundary effect, the self-weighted function $w_t$ is still needed in general. This is because
 $w_t$ is used to make sure Lemma \ref{l1} in Appendix B always holds
and so the result (\ref{joint_5_1}) is valid. Clearly,
if either (i) $Ey_{t}^{2}<\infty$ or (ii)
there exists a positive volatility coefficient $\alpha_i$ for any $i$ such that $\phi_i\not=0$,
Lemma \ref{l1} holds without $w_t$, and consequently,  $w_t$ is not needed for $Q_M$.
Otherwise, the use of $w_t$ seems necessary.



\section{Simulation studies}\label{simulation}
In this section, Monte Carlo experiments are conducted to assess the finite-sample performance of
our  Wald ($W_n$), LM ($L_n$) and QLR ($Q_{n}$) tests.


We generate 1000 replications of sample size $n=1000$ and
5000 from the following three different data generating procedures (DGPs):
\begin{flalign*}   
&\text{DGP 1}: y_t=\phi_1y_{t-1}+\eta_t\sqrt{\omega+\alpha y_{t-1}^2+ky_{t-2}^2};\\
&\text{DGP 2}: y_t=\phi_1y_{t-1}+\phi_2y_{t-2}+\eta_t\sqrt{\omega+\alpha y_{t-1}^2+ky_{t-2}^2+ky_{t-3}^2}; \\
&\text{DGP 3}: y_t=\phi_1y_{t-1}+\phi_2y_{t-2}+\eta_t\sqrt{\omega+\alpha y_{t-2}^2+ky_{t-3}^2},
\end{flalign*}
where $\phi_1=0.5$, $\phi_2=-0.3$, $\omega=1$, $\alpha=0.1$ or $0.6$, and $k=h/\sqrt{n}$ with $h\in\{0, 1, \cdots, 10\}$.
Here, three different distributions of the innovation $\eta_t$ are considered:
\begin{itemize}
	\item $\eta_t\sim \mathcal{N}(0,1)$;
	\item $\eta_t\sim\mathrm{st}_{10}$, where $\mathrm{st}_{\nu}$ is the standardized student-$t$ distribution with density
	\[
	g(x)=\frac{\Gamma[(\nu+1)/2]}{\Gamma(\nu/2)\sqrt{(\nu-2)\pi}}
\Big(1+\frac{x^2}{\nu-2}\Big)^{-(\nu+1)/2};
	\]
	\item $\eta_t\sim\mathrm{sst}_{10,2}$, where $\mathrm{sst}_{\nu,\xi}$ is the standardized skewed student-$t$ distribution in \cite{fernandez1998} with the density
\begin{eqnarray*}
f(x)=
\begin{cases}
\frac{2\rho}{\xi+1/\xi} g(\xi(\rho x+\bar{\omega})), &\mbox{if}\quad x<-\bar{\omega}/\rho, \cr \frac{2\rho}{\xi+1/\xi} g((\rho x+\bar{\omega})/\xi), &\mbox{if}\quad x\geq\bar{\omega}/\rho,
\end{cases}
\end{eqnarray*}
where
\[
\bar{\omega}=\frac{\sqrt{\nu-2}\,\Gamma[(\nu-1)/2]}{\sqrt{\pi}\,\Gamma(\nu/2)}(\xi-\xi^{-1})\,\,\,\mbox{ and }\,\,\, \rho^2=(\xi^2+\xi^{-2}-1)-\bar{\omega}^2.
\]
\end{itemize}

For each replication, we apply $W_n$, $L_n$ and $Q_{n}$ tests for the hypotheses:
\begin{equation}\label{simuhypothesis}
H_0:k=0 \quad \text{v.s.} \quad H_a:k>0.
\end{equation}
Under $H_0$ in (\ref{simuhypothesis}), there is only one coefficient on the boundary of parameter space in DGP 1, and hence the results in Theorem \ref{one coefficient} are directly applicable, while
Algorithm \ref{alg1} is needed to
calculate the critical values of $W_n$ and $Q_n$ in GDPs 2-3, since  there exist two coefficients on the boundary of parameter space in DGP 2, and a known nuisance coefficient with respect to $\alpha_1$ exists in DGP 3.
To implement Algorithm \ref{alg1}, we take $N=50,000$ in all calculations.

To justify the necessity of the self-weighting technique, all three tests are constructed based on either the GQMLE (i.e.,  $w_{t}=1$) or the S-GQMLE, where $w_{t}$ is selected as in  (\ref{weight3}) with $m=2, 3$ and $3$ for DGP 1, DGP 2 and DGP 3, respectively, and the values of $\alpha$ are chosen according to two moment situations on $y_{t}$:
\begin{itemize}
\item Case I: \: $Ey_t^6<\infty$ (with respect to $\alpha=0.1$);
\item Case II: $Ey_t^6=\infty$ and $Ey_t^2<\infty$ (with respect to $\alpha=0.6$).
\end{itemize}
Under Case I, all three tests are applicable no matter whether the self-weighted function $w_{t}$ is used, while under Case II,
all three tests may have poor performance without using $w_{t}$.

\subsection{Size comparison}\label{sizecomparison}
We first examine the size performance
of $W_{n}$, $L_{n}$ and $Q_{n}$ in finite samples at the level $\beta= 1\%$, 5\% and 10\%.

Table \ref{tablesize1} reports the sizes of all three tests in DGP1. From this table, we can see that
under Case I, (i) $W_{n}$ based on the GQMLE is always under-sized even when $n=5,000$, and $W_{n}$ based on the S-GQMLE has
a much more accurate size; (ii) both $L_{n}$ and $Q_{n}$ have a satisfactory size performance based on either the GQMLE or the S-GQMLE.
In comparison, under Case II, all three tests based on the GQMLE are seriously under-sized, while their sizes are much more accurate
based on the S-GQMLE.

\begin{table}[h]\scriptsize
	\centering
	\caption{Sizes of all three tests in DGP 1}\label{tablesize1}
 \renewcommand{\arraystretch}{1}
	\begin{tabular}{lcllllllllllllllll}
	\hline
	\multicolumn{1}{c}{$n$} & \multicolumn{1}{c}{Tests} & & &\multicolumn{6}{c}{Case I: $\alpha=0.1$}                                                                                                                                                                &  & \multicolumn{7}{c}{Case II: $\alpha=0.6$}                                                                                                                                         \\
\cline{1-1} \cline{2-2} \cline{4-10} \cline{12-18}
	\multicolumn{1}{c}{}     &   &                   & \multicolumn{3}{c}{with weight}                                                   &  & \multicolumn{3}{c}{no weight}                                                     &  & \multicolumn{3}{c}{with weight}                                                   &  & \multicolumn{3}{c}{no weight}                                                     \\
\cline{4-6} \cline{8-10} \cline{12-14} \cline{16-18}
	\multicolumn{1}{c}{}     &      &           & \multicolumn{1}{c}{1\%}  & \multicolumn{1}{c}{5\%}  & \multicolumn{1}{c}{10\%}  &  & \multicolumn{1}{c}{1\%}  & \multicolumn{1}{c}{5\%}  & \multicolumn{1}{c}{10\%}  &  & \multicolumn{1}{c}{1\%}  & \multicolumn{1}{c}{5\%}  & \multicolumn{1}{c}{10\%}  &  & \multicolumn{1}{c}{1\%}  & \multicolumn{1}{c}{5\%}  & \multicolumn{1}{c}{10\%}  \\
\cline{4-6} \cline{8-10} \cline{12-14} \cline{16-18}
	\multicolumn{1}{c}{}     & \multicolumn{1}{l}{} & &\multicolumn{15}{l}{$\eta_t\sim \mathcal{N}(0,1)$}                                                                                                                                                                                                                                                                                                       \\
\cline{4-18}
	\multicolumn{1}{c}{1000} &   $W_{n}$    &              & \multicolumn{1}{c}{0.009} & \multicolumn{1}{c}{0.048} & \multicolumn{1}{c}{0.105} &  & \multicolumn{1}{c}{0.003} & \multicolumn{1}{c}{0.038} & \multicolumn{1}{c}{0.077} &  & \multicolumn{1}{c}{0.011} & \multicolumn{1}{c}{0.051} & \multicolumn{1}{c}{0.100} &  & \multicolumn{1}{c}{0.001} & \multicolumn{1}{c}{0.013} & \multicolumn{1}{c}{0.042} \\
	\multicolumn{1}{c}{}  &   $L_{n}$     &             & \multicolumn{1}{c}{0.010} & \multicolumn{1}{c}{0.047} & \multicolumn{1}{c}{0.102} &  & \multicolumn{1}{c}{0.010} & \multicolumn{1}{c}{0.052} & \multicolumn{1}{c}{0.094} &  & \multicolumn{1}{c}{0.011} & \multicolumn{1}{c}{0.047} & \multicolumn{1}{c}{0.098} &  & \multicolumn{1}{c}{0.017} & \multicolumn{1}{c}{0.042} & \multicolumn{1}{c}{0.079} \\
	\multicolumn{1}{c}{}     & $Q_{n}$       &             & \multicolumn{1}{c}{0.011} & \multicolumn{1}{c}{0.051} & \multicolumn{1}{c}{0.105} &  & \multicolumn{1}{c}{0.008} & \multicolumn{1}{c}{0.051} & \multicolumn{1}{c}{0.087} &  & \multicolumn{1}{c}{0.012} & \multicolumn{1}{c}{0.057} & \multicolumn{1}{c}{0.098} &  & \multicolumn{1}{c}{0.005} & \multicolumn{1}{c}{0.037} & \multicolumn{1}{c}{0.076} \\
	\multicolumn{1}{c}{5000} & $W_{n}$      &              & \multicolumn{1}{c}{0.008} & \multicolumn{1}{c}{0.051} & \multicolumn{1}{c}{0.101} &  & \multicolumn{1}{c}{0.008} & \multicolumn{1}{c}{0.033} & \multicolumn{1}{c}{0.083} &  & \multicolumn{1}{c}{0.010} & \multicolumn{1}{c}{0.053} & \multicolumn{1}{c}{0.100} &  & \multicolumn{1}{c}{0.002} & \multicolumn{1}{c}{0.025} & \multicolumn{1}{c}{0.057} \\
	\multicolumn{1}{c}{}     & $L_{n}$    &                & \multicolumn{1}{c}{0.008} & \multicolumn{1}{c}{0.050} & \multicolumn{1}{c}{0.103} &  & \multicolumn{1}{c}{0.011} & \multicolumn{1}{c}{0.044} & \multicolumn{1}{c}{0.087} &  & \multicolumn{1}{c}{0.008} & \multicolumn{1}{c}{0.049} & \multicolumn{1}{c}{0.107} &  & \multicolumn{1}{c}{0.020} & \multicolumn{1}{c}{0.046} & \multicolumn{1}{c}{0.085} \\
	\multicolumn{1}{c}{}     & $Q_{n}$     &               & \multicolumn{1}{c}{0.009} & \multicolumn{1}{c}{0.051} & \multicolumn{1}{c}{0.101} &  & \multicolumn{1}{c}{0.010} & \multicolumn{1}{c}{0.037} & \multicolumn{1}{c}{0.090} &  & \multicolumn{1}{c}{0.009} & \multicolumn{1}{c}{0.051} & \multicolumn{1}{c}{0.100} &  & \multicolumn{1}{c}{0.012} & \multicolumn{1}{c}{0.044} & \multicolumn{1}{c}{0.097} \\
\cline{4-18}
	\multicolumn{1}{c}{}     & \multicolumn{1}{l}{} & &\multicolumn{15}{l}{$\eta_t\sim \mathrm{st}_{10}$}                                                                                                                                                                                                                                                                                                                \\
\cline{4-18}
	1000                     & $W_{n}$    &                & 0.010                     & 0.062                     & 0.111                     &  & 0.004                     & 0.031                     & 0.068                     &  & 0.008                     & 0.061                     & 0.119                     &  & 0.002                     & 0.021                     & 0.049                     \\
	& $L_{n}$    &                & 0.008                     & 0.053                     & 0.111                     &  & 0.013                     & 0.048                     & 0.091                     &  & 0.011                     & 0.050                     & 0.104                     &  & 0.019                     & 0.040                     & 0.064                     \\
	& $Q_{n}$  &                  & 0.011                     & 0.061                     & 0.116                     &  & 0.010                     & 0.046                     & 0.094                     &  & 0.010                     & 0.066                     & 0.123                     &  & 0.011                     & 0.043                     & 0.087                     \\
	5000                     & $W_{n}$      &              & 0.010                     & 0.055                     & 0.110                     &  & 0.005                     & 0.038                     & 0.086                     &  & 0.010                     & 0.052                     & 0.098                     &  & 0.003                     & 0.021                     & 0.052                     \\
	& $L_{n}$     &               & 0.010                     & 0.058                     & 0.104                     &  & 0.014                     & 0.055                     & 0.106                     &  & 0.010                     & 0.055                     & 0.109                     &  & 0.020                     & 0.044                     & 0.074                     \\
	& $Q_{n}$      &              & 0.011                     & 0.059                     & 0.110                     &  & 0.012                     & 0.054                     & 0.100                     &  & 0.010                     & 0.053                     & 0.099                     &  & 0.007                     & 0.046                     & 0.084                     \\
\cline{4-18}
	\multicolumn{1}{c}{}     & \multicolumn{1}{l}{} & &\multicolumn{15}{l}{$\eta_t\sim \mathrm{sst}_{10,2}$}                                                                                                                                                                                                                                                                                                           \\ \cline{4-18}
	1000                     & $W_{n}$         &          & 0.010                     & 0.049                     & 0.107                     &  & 0.002                     & 0.031                     & 0.073                     &  & 0.012                     & 0.062                     & 0.105                     &  & 0.003                     & 0.024                     & 0.053                     \\
	& $L_{n}$    &                & 0.012                     & 0.048                     & 0.101                     &  & 0.015                     & 0.039                     & 0.067                     &  & 0.013                     & 0.051                     & 0.096                     &  & 0.022                     & 0.042                     & 0.064                     \\
	& $Q_{n}$     &               & 0.013                     & 0.054                     & 0.111                     &  & 0.013                     & 0.052                     & 0.091                     &  & 0.016                     & 0.068                     & 0.113                     &  & 0.018                     & 0.050                     & 0.086                     \\
	5000                     & $W_{n}$      &              & 0.011                     & 0.059                     & 0.110                     &  & 0.006                     & 0.033                     & 0.077                     &  & 0.013                     & 0.059                     & 0.114                     &  & 0.004                     & 0.024                     & 0.056                     \\
	& $L_{n}$       &             & 0.011                     & 0.054                     & 0.112                     &  & 0.014                     & 0.044                     & 0.089                     &  & 0.012                     & 0.050                     & 0.102                     &  & 0.023                     & 0.043                     & 0.064                     \\
	& $Q_{n}$      &              & 0.013                     & 0.061                     & 0.111                     &  & 0.014                     & 0.054                     & 0.092                     &  & 0.014                     & 0.060                     & 0.113                     &  & 0.016                     & 0.049                     & 0.089                     \\ \hline
	& \multicolumn{1}{l}{} &         &                  &                           &                           &  &                           &                           &                           &  &                           &                           &                           &  &                           &                           &
\end{tabular}
\end{table}

Table \ref{tablesize2} reports the sizes of all three tests in DGP 2. From this table,
 we have similar findings as those in Table \ref{tablesize1}, except that (i) the sizes of
 $W_n$ and $L_n$ based on the GQMLE in the cases of $\eta_{t}\sim \mbox{st}_{10}$ and $\mbox{sst}_{10,2}$ are worse
 than those in the case of $\eta_{t}\sim \mathcal{N}(0, 1)$; (ii) both $W_{n}$ and $Q_{n}$
 based on the S-GQMLE are slightly oversized, especially when $n$ is small.

\begin{table}[!h]\scriptsize
	\centering
	\caption{Sizes of all three tests in DGP 2}
	\label{tablesize2}
\renewcommand{\arraystretch}{1}
	\begin{tabular}{lcllllllllllllllll}
	\hline
	\multicolumn{1}{c}{$n$} & \multicolumn{1}{c}{Tests} & & &\multicolumn{6}{c}{Case I: $\alpha=0.1$}                                                                                                                                                                &  & \multicolumn{7}{c}{Case II: $\alpha=0.6$}                                                                                                                                         \\
\cline{1-1} \cline{2-2} \cline{4-10} \cline{12-18}
	\multicolumn{1}{c}{}     &   &                   & \multicolumn{3}{c}{with weight}                                                   &  & \multicolumn{3}{c}{no weight}                                                     &  & \multicolumn{3}{c}{with weight}                                                   &  & \multicolumn{3}{c}{no weight}                                                     \\
\cline{4-6} \cline{8-10} \cline{12-14} \cline{16-18}
	\multicolumn{1}{c}{}     &      &           & \multicolumn{1}{c}{1\%}  & \multicolumn{1}{c}{5\%}  & \multicolumn{1}{c}{10\%}  &  & \multicolumn{1}{c}{1\%}  & \multicolumn{1}{c}{5\%}  & \multicolumn{1}{c}{10\%}  &  & \multicolumn{1}{c}{1\%}  & \multicolumn{1}{c}{5\%}  & \multicolumn{1}{c}{10\%}  &  & \multicolumn{1}{c}{1\%}  & \multicolumn{1}{c}{5\%}  & \multicolumn{1}{c}{10\%}  \\
\cline{4-6} \cline{8-10} \cline{12-14} \cline{16-18}
	\multicolumn{1}{c}{}     & \multicolumn{1}{l}{} & &\multicolumn{15}{l}{$\eta_t\sim \mathcal{N}(0,1)$}                                                                                                                                                                                                                                                                                                       \\
\cline{4-18}
		\multicolumn{1}{c}{1000} & $W_{n}$        &            & \multicolumn{1}{c}{0.016} & \multicolumn{1}{c}{0.065} & \multicolumn{1}{c}{0.118} &  & \multicolumn{1}{c}{0.001} & \multicolumn{1}{c}{0.022} & \multicolumn{1}{c}{0.055} &  & \multicolumn{1}{c}{0.017} & \multicolumn{1}{c}{0.075} & \multicolumn{1}{c}{0.129} &  & \multicolumn{1}{c}{0.000}  & \multicolumn{1}{c}{0.010} & \multicolumn{1}{c}{0.026} \\
		\multicolumn{1}{c}{}     & $L_{n}$        &                     & \multicolumn{1}{c}{0.017} & \multicolumn{1}{c}{0.056} & \multicolumn{1}{c}{0.119} &  & \multicolumn{1}{c}{0.011} & \multicolumn{1}{c}{0.037} & \multicolumn{1}{c}{0.090} &  & \multicolumn{1}{c}{0.011} & \multicolumn{1}{c}{0.054} & \multicolumn{1}{c}{0.101} &  & \multicolumn{1}{c}{0.019}  & \multicolumn{1}{c}{0.048} & \multicolumn{1}{c}{0.075} \\
		\multicolumn{1}{c}{}     & $Q_{n}$        &                     & \multicolumn{1}{c}{0.017} & \multicolumn{1}{c}{0.070} & \multicolumn{1}{c}{0.127} &  & \multicolumn{1}{c}{0.010} & \multicolumn{1}{c}{0.036} & \multicolumn{1}{c}{0.082} &  & \multicolumn{1}{c}{0.017} & \multicolumn{1}{c}{0.074} & \multicolumn{1}{c}{0.133} &  & \multicolumn{1}{c}{0.005}  & \multicolumn{1}{c}{0.031} & \multicolumn{1}{c}{0.066} \\
		\multicolumn{1}{c}{5000} & $W_{n}$        &                     & \multicolumn{1}{c}{0.013} & \multicolumn{1}{c}{0.055} & \multicolumn{1}{c}{0.109} &  & \multicolumn{1}{c}{0.005} & \multicolumn{1}{c}{0.031} & \multicolumn{1}{c}{0.074} &  & \multicolumn{1}{c}{0.015} & \multicolumn{1}{c}{0.060} & \multicolumn{1}{c}{0.111} &  & \multicolumn{1}{c}{0.001}  & \multicolumn{1}{c}{0.017} & \multicolumn{1}{c}{0.052} \\
		\multicolumn{1}{c}{}     & $L_{n}$        &                     & \multicolumn{1}{c}{0.012} & \multicolumn{1}{c}{0.051} & \multicolumn{1}{c}{0.096} &  & \multicolumn{1}{c}{0.016} & \multicolumn{1}{c}{0.056} & \multicolumn{1}{c}{0.106} &  & \multicolumn{1}{c}{0.009} & \multicolumn{1}{c}{0.052} & \multicolumn{1}{c}{0.102} &  & \multicolumn{1}{c}{0.023}  & \multicolumn{1}{c}{0.057} & \multicolumn{1}{c}{0.099} \\
		\multicolumn{1}{c}{}     & $Q_{n}$        &                     & \multicolumn{1}{c}{0.015} & \multicolumn{1}{c}{0.054} & \multicolumn{1}{c}{0.113} &  & \multicolumn{1}{c}{0.011} & \multicolumn{1}{c}{0.047} & \multicolumn{1}{c}{0.092} &  & \multicolumn{1}{c}{0.014} & \multicolumn{1}{c}{0.065} & \multicolumn{1}{c}{0.111} &  & \multicolumn{1}{c}{0.008}  & \multicolumn{1}{c}{0.045} & \multicolumn{1}{c}{0.091} \\ \cline{4-18}
		\multicolumn{1}{c}{}     & \multicolumn{1}{l}{} & &\multicolumn{15}{l}{$\eta_t\sim \mathrm{st}_{10}$}                                                                                                                                                                                                                                                                                                                 \\ \cline{4-18}
		1000                     & $W_{n}$        &                    & 0.016                     & 0.066                     & 0.121                     &  & 0.004                     & 0.026                     & 0.062                     &  & 0.019                     & 0.073                     & 0.132                     &  & 0.002                      & 0.013                     & 0.035                     \\
		& $L_{n}$        &                     & 0.014                     & 0.050                     & 0.112                     &  & 0.022                     & 0.054                     & 0.100                     &  & 0.015                     & 0.053                     & 0.106                     &  & 0.025                      & 0.053                     & 0.075                     \\
		& $Q_{n}$        &                     & 0.019                     & 0.073                     & 0.136                     &  & 0.013                     & 0.053                     & 0.089                     &  & 0.024                     & 0.084                     & 0.145                     &  & 0.014                      & 0.040                     & 0.071                     \\
		5000                     & $W_{n}$        &                     & 0.011                     & 0.054                     & 0.101                     &  & 0.003                     & 0.028                     & 0.065                     &  & 0.016                     & 0.057                     & 0.109                     &  & 0.002                      & 0.019                     & 0.044                     \\
		& $L_{n}$        &                     & 0.009                     & 0.052                     & 0.100                     &  & 0.010                     & 0.040                     & 0.086                     &  & 0.010                     & 0.051                     & 0.103                     &  & 0.022                      & 0.050                     & 0.082                     \\
		& $Q_{n}$        &                     & 0.011                     & 0.058                     & 0.106                     &  & 0.009                     & 0.044                     & 0.087                     &  & 0.017                     & 0.060                     & 0.112                     &  & 0.011                      & 0.042                     & 0.086                     \\
\cline{4-18}
		\multicolumn{1}{c}{}     & \multicolumn{1}{l}{} & & \multicolumn{15}{l}{$\eta_t\sim \mathrm{sst}_{10,2}$}                                                                                                                                                                                                                                                                                                            \\ \cline{4-18}
		1000                     & $W_{n}$        &                     & 0.022                     & 0.076                     & 0.139                     &  & 0.001                     & 0.022                     & 0.048                     &  & 0.021                     & 0.084                     & 0.154                     &  & 0.002                      & 0.014                     & 0.043                     \\
		& $L_{n}$        &                     & 0.017                     & 0.061                     & 0.109                     &  & 0.026                     & 0.058                     & 0.080                     &  & 0.016                     & 0.056                     & 0.116                     &  & 0.021                      & 0.046                     & 0.062                     \\
		& $Q_{n}$        &                     & 0.032                     & 0.089                     & 0.146                     &  & 0.019                     & 0.051                     & 0.084                     &  & 0.031                     & 0.098                     & 0.167                     &  & 0.014                      & 0.044                     & 0.083                     \\
		5000                     & $W_{n}$        &                     & 0.015                     & 0.061                     & 0.111                     &  & 0.002                     & 0.022                     & 0.060                     &  & 0.012                     & 0.066                     & 0.126                     &  & 0.006                      & 0.019                     & 0.052                     \\
		& $L_{n}$        &                     & 0.011                     & 0.053                     & 0.108                     &  & 0.025                     & 0.053                     & 0.086                     &  & 0.009                     & 0.050                     & 0.102                     &  & 0.031                      & 0.051                     & 0.070                     \\
		& $Q_{n}$        &                     & 0.018                     & 0.061                     & 0.113                     &  & 0.017                     & 0.048                     & 0.092                     &  & 0.013                     & 0.070                     & 0.128                     &  & 0.017                      & 0.049                     & 0.092                     \\ \hline
		& \multicolumn{1}{l}{} &     &                      &                           &                           &  &                           &                           &                           &  &                           &                           &                           &  &                            &                           &
	\end{tabular}
\end{table}

 Table \ref{tablesize3} reports the sizes of all three tests in DGP 3. In this case, we can reach the similar conclusions as
in Table \ref{tablesize1}, except that all tests based on the GQMLE under Case II are undersized even when $n$ is large.

\begin{table}[!h]\scriptsize
	\centering
	\caption{Sizes of all three tests in DGP 3}
	\label{tablesize3}
\renewcommand{\arraystretch}{1}
	\begin{tabular}{lcllllllllllllllll}
		\hline
	\multicolumn{1}{c}{$n$} & \multicolumn{1}{c}{Tests} & & &\multicolumn{6}{c}{Case I: $\alpha=0.1$}                                                                                                                                                                &  & \multicolumn{7}{c}{Case II: $\alpha=0.6$}                                                                                                                                         \\
\cline{1-1} \cline{2-2} \cline{4-10} \cline{12-18}
	\multicolumn{1}{c}{}     &   &                   & \multicolumn{3}{c}{with weight}                                                   &  & \multicolumn{3}{c}{no weight}                                                     &  & \multicolumn{3}{c}{with weight}                                                   &  & \multicolumn{3}{c}{no weight}                                                     \\
\cline{4-6} \cline{8-10} \cline{12-14} \cline{16-18}
	\multicolumn{1}{c}{}     &      &           & \multicolumn{1}{c}{1\%}  & \multicolumn{1}{c}{5\%}  & \multicolumn{1}{c}{10\%}  &  & \multicolumn{1}{c}{1\%}  & \multicolumn{1}{c}{5\%}  & \multicolumn{1}{c}{10\%}  &  & \multicolumn{1}{c}{1\%}  & \multicolumn{1}{c}{5\%}  & \multicolumn{1}{c}{10\%}  &  & \multicolumn{1}{c}{1\%}  & \multicolumn{1}{c}{5\%}  & \multicolumn{1}{c}{10\%}  \\
\cline{4-6} \cline{8-10} \cline{12-14} \cline{16-18}
	\multicolumn{1}{c}{}     & \multicolumn{1}{l}{} & &\multicolumn{15}{l}{$\eta_t\sim \mathcal{N}(0,1)$}                                                                                                                                                                                                                                                                                                       \\
\cline{4-18}
		\multicolumn{1}{c}{1000} & $W_{n}$  &                  & \multicolumn{1}{c}{0.016} & \multicolumn{1}{c}{0.060} & \multicolumn{1}{c}{0.112} &  & \multicolumn{1}{c}{0.007} & \multicolumn{1}{c}{0.034} & \multicolumn{1}{c}{0.076} &  & \multicolumn{1}{c}{0.016} & \multicolumn{1}{c}{0.071} & \multicolumn{1}{c}{0.135} &  & \multicolumn{1}{c}{0.002} & \multicolumn{1}{c}{0.037} & \multicolumn{1}{c}{0.049} \\
		\multicolumn{1}{c}{}     & $L_{n}$  &                    & \multicolumn{1}{c}{0.018} & \multicolumn{1}{c}{0.061} & \multicolumn{1}{c}{0.116} &  & \multicolumn{1}{c}{0.013} & \multicolumn{1}{c}{0.050} & \multicolumn{1}{c}{0.102} &  & \multicolumn{1}{c}{0.009} & \multicolumn{1}{c}{0.043} & \multicolumn{1}{c}{0.089} &  & \multicolumn{1}{c}{0.014} & \multicolumn{1}{c}{0.034} & \multicolumn{1}{c}{0.077} \\
		\multicolumn{1}{c}{}     & $Q_{n}$  &                    & \multicolumn{1}{c}{0.018} & \multicolumn{1}{c}{0.063} & \multicolumn{1}{c}{0.114} &  & \multicolumn{1}{c}{0.013} & \multicolumn{1}{c}{0.045} & \multicolumn{1}{c}{0.098} &  & \multicolumn{1}{c}{0.015} & \multicolumn{1}{c}{0.064} & \multicolumn{1}{c}{0.120} &  & \multicolumn{1}{c}{0.006} & \multicolumn{1}{c}{0.050} & \multicolumn{1}{c}{0.065} \\
		\multicolumn{1}{c}{5000} & $W_{n}$  &                    & \multicolumn{1}{c}{0.011} & \multicolumn{1}{c}{0.056} & \multicolumn{1}{c}{0.104} &  & \multicolumn{1}{c}{0.006} & \multicolumn{1}{c}{0.036} & \multicolumn{1}{c}{0.084} &  & \multicolumn{1}{c}{0.015} & \multicolumn{1}{c}{0.077} & \multicolumn{1}{c}{0.131} &  & \multicolumn{1}{c}{0.002} & \multicolumn{1}{c}{0.024} & \multicolumn{1}{c}{0.063} \\
		\multicolumn{1}{c}{}     & $L_{n}$  &                    & \multicolumn{1}{c}{0.010} & \multicolumn{1}{c}{0.051} & \multicolumn{1}{c}{0.102} &  & \multicolumn{1}{c}{0.012} & \multicolumn{1}{c}{0.046} & \multicolumn{1}{c}{0.094} &  & \multicolumn{1}{c}{0.009} & \multicolumn{1}{c}{0.046} & \multicolumn{1}{c}{0.105} &  & \multicolumn{1}{c}{0.014} & \multicolumn{1}{c}{0.041} & \multicolumn{1}{c}{0.077} \\
		\multicolumn{1}{c}{}     & $Q_{n}$  &                    & \multicolumn{1}{c}{0.012} & \multicolumn{1}{c}{0.056} & \multicolumn{1}{c}{0.105} &  & \multicolumn{1}{c}{0.009} & \multicolumn{1}{c}{0.045} & \multicolumn{1}{c}{0.096} &  & \multicolumn{1}{c}{0.014} & \multicolumn{1}{c}{0.071} & \multicolumn{1}{c}{0.120} &  & \multicolumn{1}{c}{0.005} & \multicolumn{1}{c}{0.035} & \multicolumn{1}{c}{0.079} \\
\cline{4-18}
		\multicolumn{1}{c}{}     & \multicolumn{1}{l}{} & &\multicolumn{15}{l}{$\eta_t\sim \mathrm{st}_{10}$}                                                                                                                                                                                                                                                                                                                \\ \cline{4-18}
		1000                     & $W_{n}$  &                    & 0.010                     & 0.057                     & 0.119                     &  & 0.002                     & 0.037                     & 0.089                     &  & 0.013                     & 0.074                     & 0.136                     &  & 0.003                     & 0.024                     & 0.054                     \\
		& $L_{n}$  &                    & 0.013                     & 0.060                     & 0.118                     &  & 0.010                     & 0.054                     & 0.103                     &  & 0.010                     & 0.052                     & 0.103                     &  & 0.021                     & 0.048                     & 0.079                     \\
		& $Q_{n}$  &                    & 0.014                     & 0.065                     & 0.125                     &  & 0.012                     & 0.058                     & 0.109                     &  & 0.015                     & 0.073                     & 0.132                     &  & 0.011                     & 0.041                     & 0.078                     \\
		5000                     & $W_{n}$  &                    & 0.010                     & 0.056                     & 0.117                     &  & 0.007                     & 0.046                     & 0.104                     &  & 0.013                     & 0.056                     & 0.116                     &  & 0.005                     & 0.028                     & 0.059                     \\
		& $L_{n}$  &                    & 0.013                     & 0.058                     & 0.114                     &  & 0.014                     & 0.058                     & 0.103                     &  & 0.010                     & 0.050                     & 0.093                     &  & 0.014                     & 0.044                     & 0.091                     \\
		& $Q_{n}$  &                    & 0.013                     & 0.061                     & 0.122                     &  & 0.013                     & 0.059                     & 0.116                     &  & 0.013                     & 0.053                     & 0.113                     &  & 0.011                     & 0.039                     & 0.071                     \\
\cline{4-18}
		\multicolumn{1}{c}{}     & \multicolumn{1}{l}{} & &\multicolumn{15}{l}{$\eta_t\sim \mathrm{sst}_{10,2}$}                                                                                                                                                                                                                                                                                                           \\ \cline{4-18}
		1000                     & $W_{n}$  &                    & 0.017                     & 0.071                     & 0.130                     &  & 0.005                     & 0.024                     & 0.068                     &  & 0.020                     & 0.080                     & 0.143                     &  & 0.035                     & 0.022                     & 0.056                     \\
		& $L_{n}$  &                    & 0.015                     & 0.065                     & 0.124                     &  & 0.014                     & 0.040                     & 0.082                     &  & 0.013                     & 0.054                     & 0.101                     &  & 0.017                     & 0.046                     & 0.071                     \\
		& $Q_{n}$  &                    & 0.021                     & 0.084                     & 0.131                     &  & 0.013                     & 0.044                     & 0.095                     &  & 0.024                     & 0.079                     & 0.139                     &  & 0.009                     & 0.038                     & 0.072                     \\
		5000                     & $W_{n}$  &                    & 0.013                     & 0.055                     & 0.109                     &  & 0.007                     & 0.036                     & 0.079                     &  & 0.011                     & 0.060                     & 0.123                     &  & 0.002                     & 0.024                     & 0.061                     \\
		& $L_{n}$  &                    & 0.013                     & 0.059                     & 0.108                     &  & 0.016                     & 0.049                     & 0.082                     &  & 0.008                     & 0.038                     & 0.089                     &  & 0.013                     & 0.040                     & 0.072                     \\
		& $Q_{n}$  &                    & 0.016                     & 0.058                     & 0.113                     &  & 0.013                     & 0.047                     & 0.095                     &  & 0.013                     & 0.058                     & 0.121                     &  & 0.007                     & 0.041                     & 0.082                     \\ \hline
		& \multicolumn{1}{l}{} &      &                     &                           &                           &  &                           &                           &                           &  &                           &                           &                           &  &                           &                           &
	\end{tabular}
\end{table}

Overall, all three tests based on the S-GQMLE have accurate sizes in all examined cases, while
their size performance based on the GQMLE is not satisfactory under Case II,
indicating the necessity of the self-weighting technique when $y_{t}$ has an infinite sixth moment.

\subsection{Power comparison}
We next compare the local power of all three tests in finite samples at the level $\beta= 5\%$.
Under Case I, the size-adjusted local power is computed to do a better power comparison, while under Case II, the size-adjusted local power is
only calculated based on the S-GQMLE, since the sizes of all three tests based on the GQMLE are not
accurate in this case.
Also, we only show the power plot for $\eta_t\sim \mathcal{N}(0,1)$, since the power plots for $\eta_t\sim \mbox{st}_{10}$ and $\eta_t\sim \mbox{sst}_{10,2}$ are similar, and they are available upon request but not depicted here for saving the space.

\begin{figure}[!h]
	\centering
    \includegraphics[width=15cm,height=12cm]{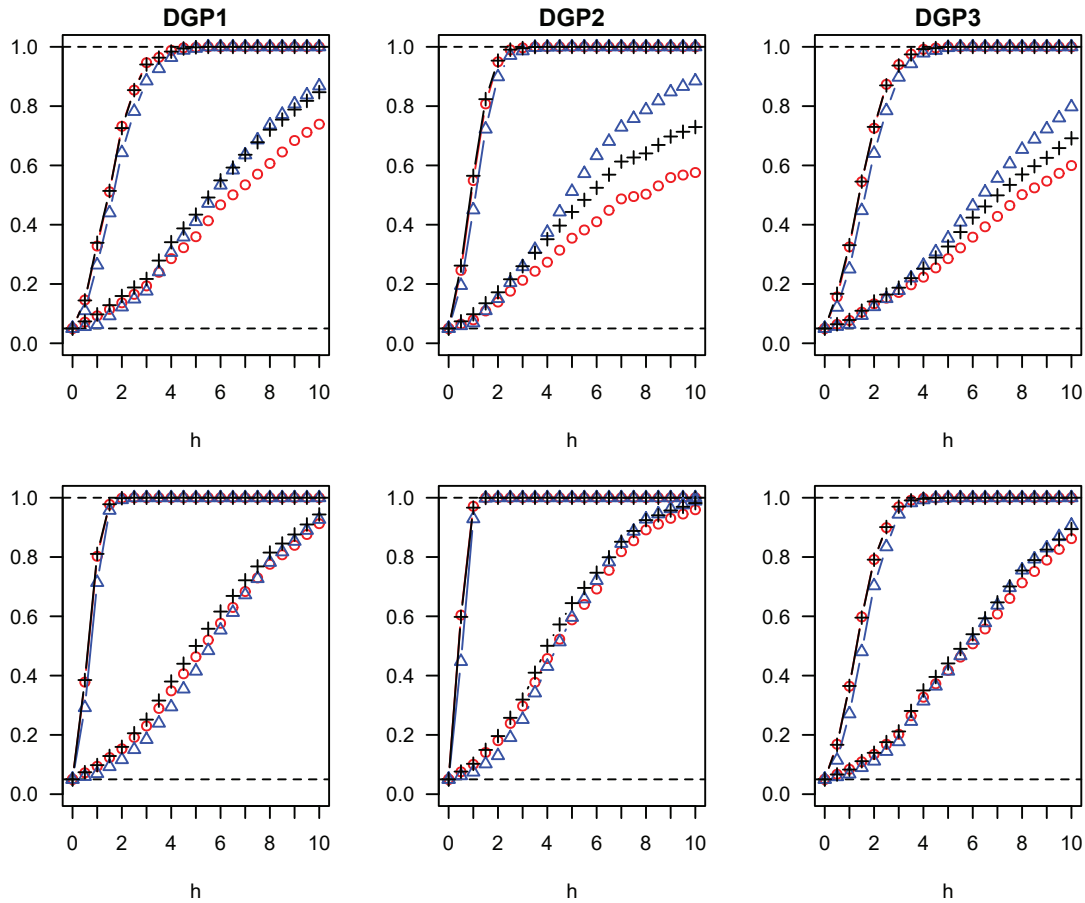}
	\caption{Local power comparison in Case I. (Upper: $n=1,000$; bottom: $n=5,000$.
		Circle $(\circ)$: $W_{n}$ ; triangle $(\triangle)$: $L_{n}$; cross $(+)$: $Q_{n}$. With weight: points; without weight: points and lines)}
	\label{alpha1}
\end{figure}

Fig \ref{alpha1} plots the size-adjusted local power (across $h$) for all three tests under Case I.
From this figure, we can find that (i)  the power of all three tests based on the GQMLE is always higher than that based on the S-GQMLE;
(ii) based on the S-GQMLE, $L_n$ is most powerful when $n$ is small (i.e., $n=1,000$), while the power advantage of
$L_{n}$ over other two tests becomes vague when $n$ is large (i.e., $n=5,000$); (iii)  all three tests are consistent in all examined cases;
(iv) when  $n=5,000$, $W_n$ and $Q_n$ are more powerful than $L_n$ in DGP 1, and this is consistent to the result in Theorem \ref{localpower}.

\begin{figure}[!h]
	\centering
    \includegraphics[width=15cm,height=12cm]{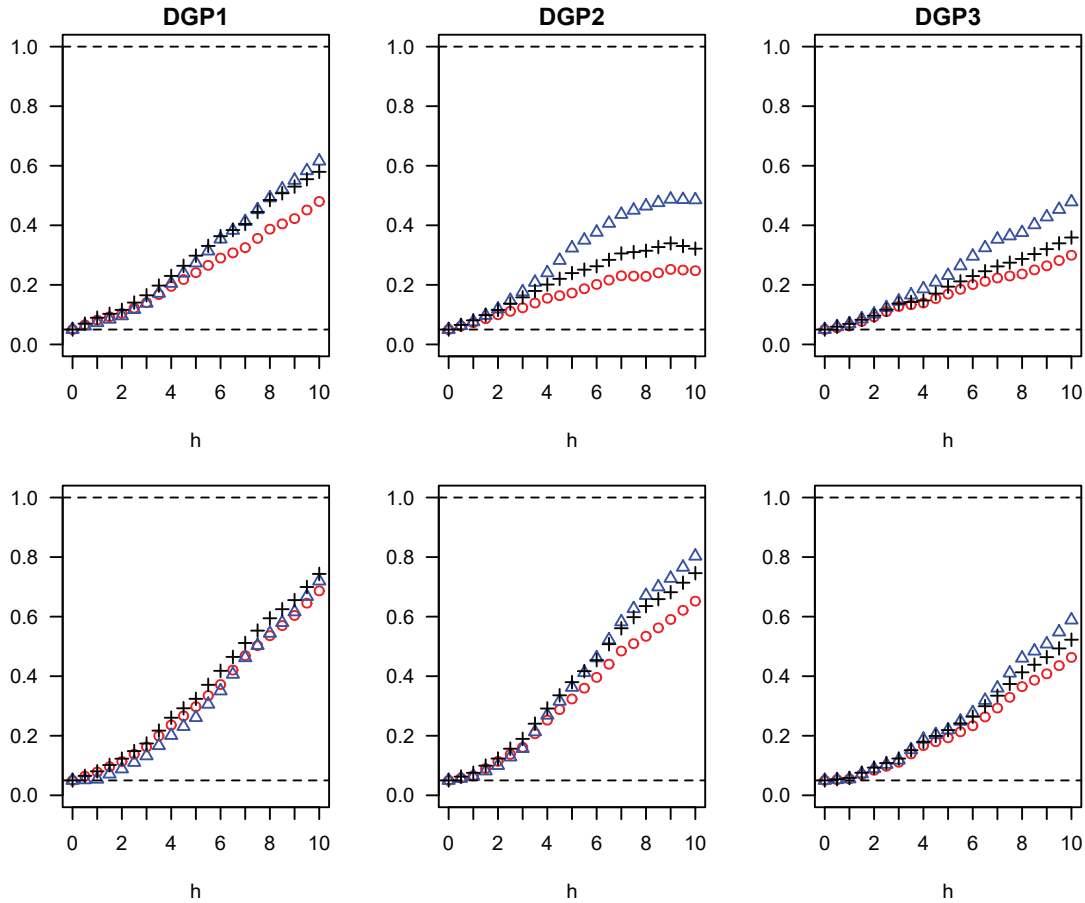}
	\caption{Local power comparison in  Case II with weight.(Upper: $n=1,000$; bottom: $n=5,000$.
	Circle $(\circ)$: $W_n$ ; triangle $(\triangle)$: $L_n$; cross $(+)$: $Q_n$)}
	\label{normselfalpha2}
\end{figure}

Fig \ref{normselfalpha2} plots the size-adjusted local power (across $h$) for all three tests under Case II.
From this figure, we have the similar findings as those in Fig \ref{alpha1},  except that (i) when $n$ is small, the local power for all three tests is relatively small even when $h$ is large; (ii) $L_n$ becomes more powerful than $W_n$ in DGP 1, especially for large $h$.

Overall, when $y_{t}$ has a finite sixth moment, the power comparison suggests us to use all three tests based on the GQMLE instead of the S-GQMLE.
On the other hand, when $y_{t}$ has an infinite sixth moment, all three tests based on the S-GQMLE have a satisfactory power.

\section{A real example}\label{emexample}
\begin{figure}[!h]
	\centering
        \includegraphics[width=8cm,height=7cm]{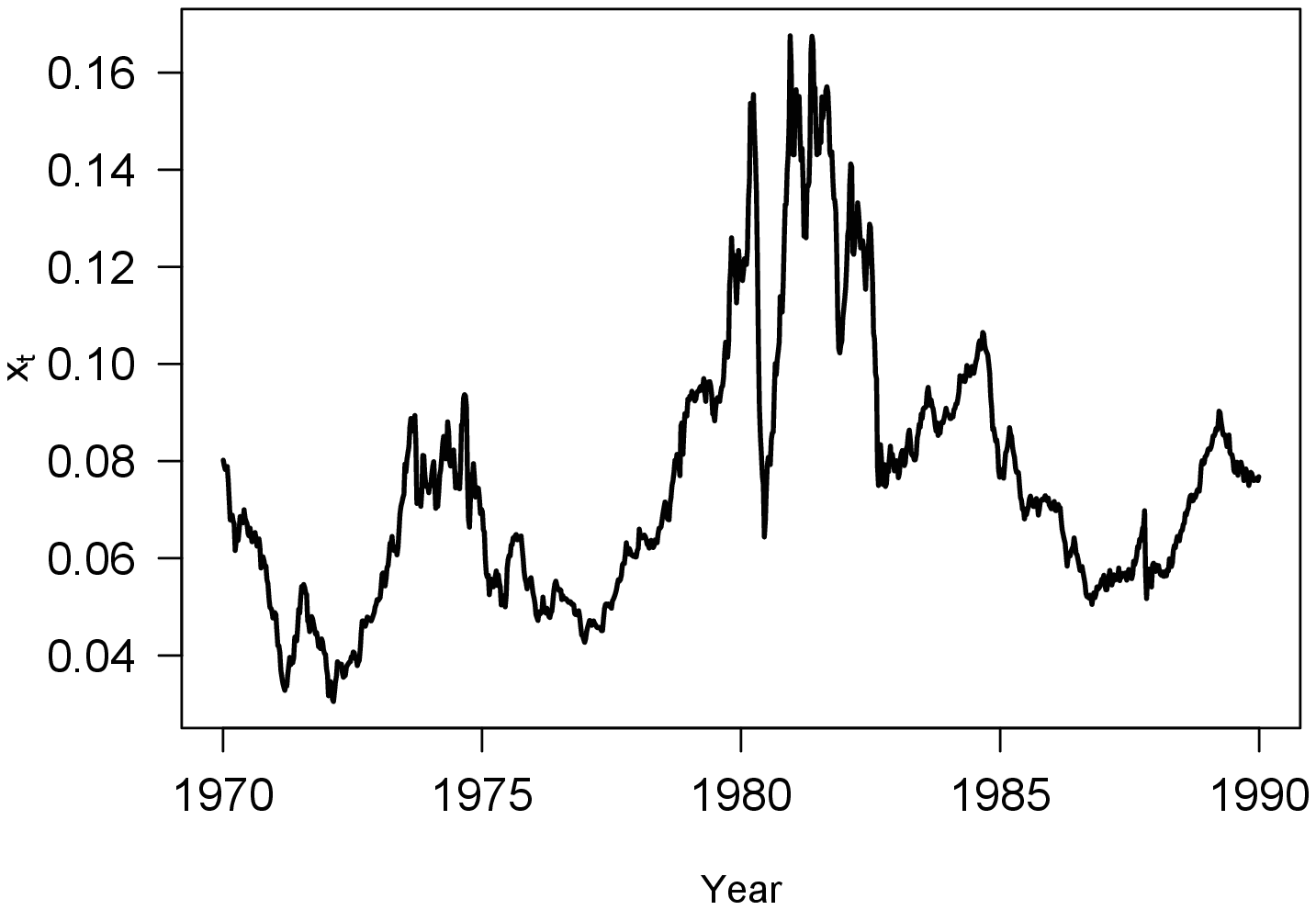}
        \includegraphics[width=8cm,height=7cm]{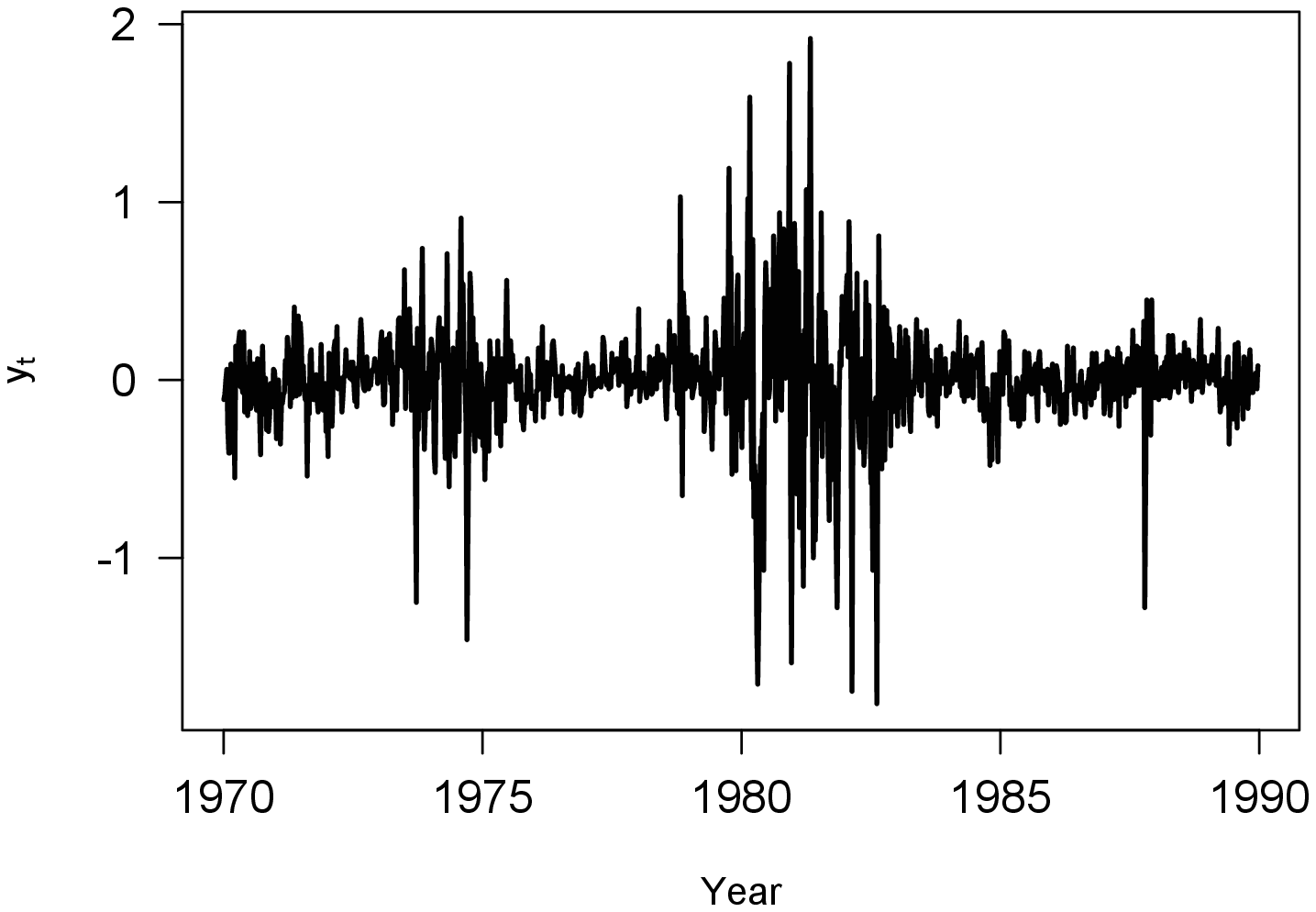}
		\caption{Plots of $\{x_t\}$ and $\{y_t\}$.}
		\label{xyseries}
\end{figure}

In this section, we study the weekly 3-Month Treasury Bill rate of second market in U.S. from January, 1970 to December, 1989, which
has 1044 observations in total. Denote this data set  by $\{x_{t}\}_{t=0}^{1043}$, and its difference (in percentage) by
$\{y_{t}\}_{t=1}^{1043}$, where  $y_t=100(x_t-x_{t-1})$, and both $\{x_{t}\}_{t=0}^{1043}$ and $\{y_{t}\}_{t=1}^{1043}$ are plotted in Fig \ref{xyseries}.
To begin with, we apply the Phillips-Perron test $Z(\hat{\alpha})$ in \cite{PP1988}
to
$\{x_{t}\}_{t=0}^{1043}$ and $\{y_{t}\}_{t=1}^{1043}$, and find that the corresponding p-values are 0.3474 and 0.0010,
respectively. These results imply that
 $y_{t}$ is stationary while $x_t$ is not. Hence, we consider $y_t$ instead of $x_t$ in the sequel.

\begin{figure}[!h]
	\centering
    \includegraphics[width=12cm,height=8cm]{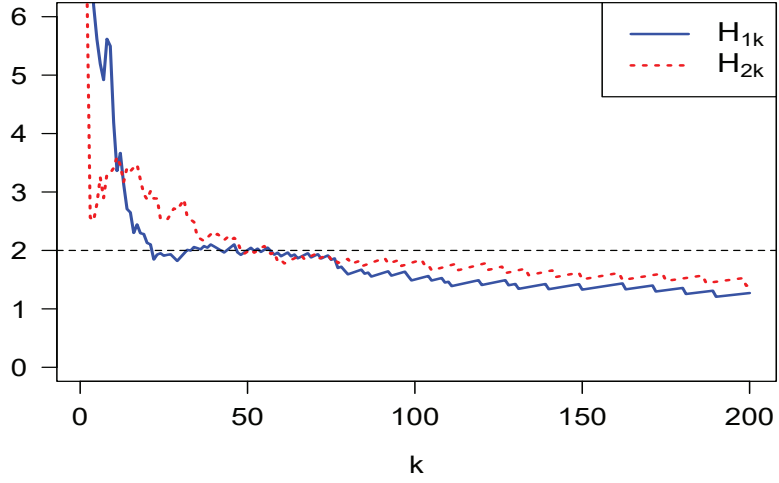}
	\caption{Right-tail Hill estimators $H_{1k} ($--$)$ and left-tail Hill estimators $H_{2k} ($- -$)$.}
	\label{hill}
\end{figure}

Next, based on $\{y_t\}_{t=1}^{1043}$, we use the Hill estimator to estimate the tail index of $y_t$.
Fig \ref{hill} plots the right-tail and left-tail Hill estimators of $y_{t}$ given by
\[
H_{1k}=\left\{ \frac{1}{k}\sum_{i=1}^{k}\log\big(\frac{y_{(i)}}{y_{(k+1)}}\big)  \right\}^{-1} \qquad \text{and} \qquad H_{2k}=\left\{ \frac{1}{k}\sum_{i=1}^{k}\log\big(\frac{y_{(n-i+1)}}{y_{(n-k)}}\big)  \right\}^{-1},
\]
respectively, where $\{y_{(t)}\}_{t=1}^{1043}$ are the ascending order statistics of $\{y_t\}_{t=1}^{1043}$. From Fig \ref{hill}, we can see that both the right-tail and left-tail indices of $y_t$ are most likely less than 2. Hence,
$y_t$ seems to be heavy-tailed with an infinite second moment, indicating the use of self-weighting technique to model $\{y_{t}\}_{t=1}^{1043}$.

Based on these facts, we fit $\{y_{t}\}_{t=1}^{1043}$ by an ADAR model:
\begin{flalign}\label{fit_model_1}
\left\{
\begin{array}{l}
y_t=-0.0002_{(-0.0137)}+0.2733_{(6.4160)}y_{t-1}-0.0097_{(0.2506)}y_{t-2}+0.0405_{(1.0887)}y_{t-3}\\
\,\,\,\,\,+0.1689_{(4.7712)}y_{t-4}
+0.0303_{(0.8535)}y_{t-5}+0.0186_{(0.4526)}y_{t-6}-0.0317_{(0.8661)}y_{t-7}+\eta_t\sqrt{h_t},\\
h_t=0.0062_{(3.6471)}+0.3560_{(3.8280)}y_{t-1}^2+0.1274_{(1.9969)}y_{t-2}^2+
0.1024_{(1.7595)}y_{t-3}^2\\
\,\,\,\,\,+0.0537_{(1.0783)}y_{t-4}^2+0.0322_{(0.6880)}y_{t-5}^2+0.2871_{(3.4466)}y_{t-6}^2+0.0791_{(1.4785)}y_{t-7}^2,
\end{array}
\right.
\end{flalign}
where model (\ref{fit_model_1}) is estimated by the S-GQMLE with the self-weighted function $w_t$ defined in (\ref{weight3}) with $m=7$, and
the corresponding values of $t_{n}$ are given in parentheses.

In the following, we use a two-step test procedure to examine the significance of coefficients in model (\ref{fit_model_1}). First, we apply
the tests $W_n$, $L_n$ and $Q_n$
to detect the null hypothesis $H_{0}^{(m)}: u_0=\phi_{20}=\phi_{30}=\phi_{50}=\phi_{60}=\phi_{70}=0$. Since $H_{0}^{(m)}$ is designed for the conditional mean coefficients, all $W_n$, $L_n$ and $Q_n$ tests have standard asymptotics, and their corresponding p-values  are  0.831, 0.829 and 0.907, respectively. Hence, we can not reject $H_{0}^{(m)}$ at the level 5\%, and then fit $\{y_{t}\}_{t=1}^{1043}$ by the following reduced ADAR model:
\begin{flalign}\label{fit_model_2}
\left\{
\begin{array}{l}
y_t=0.2699_{(6.4569)}y_{t-1}+0.1765_{(5.1912)}y_{t-4}+\eta_t\sqrt{h_t},\\
h_t=0.0063_{(3.7059)}+0.3679_{(3.9773)}y_{t-1}^2+0.1254_{(2.0096)}y_{t-2}^2+
0.1109_{(1.8990)}y_{t-3}^2\\
\,\,\,\,\,\quad+0.0480_{(1.0063)}y_{t-4}^2+0.0305_{(0.6489)}y_{t-5}^2+0.2897_{(3.5372)}y_{t-6}^2+0.0746_{(1.4458)}y_{t-7}^2,
\end{array}
\right.
\end{flalign}
where the preceding model is estimated in the same way as model (\ref{fit_model_1}). Second, since
the values of $t_{n}$ for
the coefficients $\alpha_4$ and $\alpha_5$ are relatively small in model (\ref{fit_model_2}),
we further apply the tests $W_n$, $L_n$ and $Q_n$
to detect the null hypothesis $H_{0}^{(v)'}: \alpha_{40}=\alpha_{50}=0$. In this case,  the p-values of $W_n$, $L_n$ and $Q_n$ are  0.197, 0.391 and 0.181, respectively,
where the p-value of $L_n$ is computed by Theorem \ref{testmain}(ii), and the p-values of $W_n$ and $Q_n$
are computed by Algorithm \ref{alg1}.
Here, we take $d_1=8$, $d_2=0$, $d_3=2$, $\theta_{0}^{(1)}=(\phi_{10},\phi_{40},\omega_0,\alpha_{10},\alpha_{20},\alpha_{30},\alpha_{60},\alpha_{70})'$,
$\theta_{0}^{(3)}=(\alpha_{40},\alpha_{50})'$, $\Lambda=\mathbb{R}^{d_1}\times[0,\infty]^{d_3}$, and
$\Lambda_{|3}=\mathbb{R}^{d_1}\times\{0\}^{d_3}$ in the implementation of Algorithm \ref{alg1}.
Since $H_{0}^{(v)'}$ can not be rejected at the level 5\% as indicated by all three tests,
we refit $\{y_{t}\}_{t=1}^{1043}$ by a simpler reduced ADAR model:
\begin{flalign}\label{fit_model_3}
\left\{
\begin{array}{l}
y_t=0.2704_{(6.4381)}y_{t-1}+0.1778_{(5.6444)}y_{t-4}+\eta_t\sqrt{h_t},\\
h_t=0.0069_{(4.0588)}+0.3943_{(4.2627)}y_{t-1}^2+0.1351_{(2.1790)}y_{t-2}^2+0.1326_{(2.2784)}y_{t-3}^2\\
\,\,\,\,\,\quad+0.2960_{(3.7374)}y_{t-6}^2+0.0787_{(1.5835)}y_{t-7}^2,
\end{array}
\right.
\end{flalign}
where the preceding model is estimated in the same way as model (\ref{fit_model_1}).
For model (\ref{fit_model_3}), the tests $W_n$, $L_n$ and $Q_n$ (with p-values close to zero) imply that
the null hypothesis $H_0^{(v)''}: \omega_0=\phi_{10}=\phi_{20}=\phi_{30}=\phi_{60}=\phi_{70}=0$ is rejected.
Meanwhile, by using the results in Theorem \ref{one coefficient},
the p-values of the tests $t_n$, $W_n$, $L_n$ and $Q_n$ for the null hypothesis $H_0^{(v)'''}: \phi_{70}=0$ are
0.056, 0.056, 0.005 and 0.008, respectively, indicating that $\phi_{70}$ is not zero.
These results may suggest that our presumption of $d_2=0$ above is appropriate.
Moreover, we find that
model (\ref{fit_model_3})  is adequate at the level 5\%,
since its p-values of portmanteau tests $Q_{6}$, $Q_{12}$ and $Q_{18}$ calculated
from the standard chi-squared limiting null distribution are 0.984, 0.840 and 0.760, respectively.

 As a comparison, we also fit $\{y_{t}\}_{t=1}^{1043}$ by an AR-GARCH model:
\begin{flalign}\label{fit_ar_garch}
\left\{
\begin{array}{l}
 y_t=0.2185_{(5.6901)}y_{t-1}+0.1183_{(3.1547)}y_{t-4}+\epsilon_t, \,\,\,\,\epsilon_t=\eta_t\sqrt{h_t},\\
 h_t=0.0014_{(3.5012)}+0.3039_{(6.3445)}\epsilon_{t-1}^2+0.7328_{(22.0723)}h_{t-1},
 \end{array}
 \right.
\end{flalign}
where model (\ref{fit_ar_garch}) is estimated by the S-GQMLE in \cite{ling2007self}  with the self-weighted function $w_t$ defined in (\ref{weight3}), and
the corresponding values of $t_{n}$ are given in parentheses. The values of AIC and BIC of model (\ref{fit_ar_garch})  are -1799.5 and -1774.8, respectively, while
these of model (\ref{fit_model_3}) are -1841.5 and -1801.9, respectively.
In view of this,  model (\ref{fit_model_3}) is more suitable than model (\ref{fit_ar_garch}) to fit $\{y_{t}\}_{t=1}^{1043}$.


\section*{Acknowledgements}
The authors greatly appreciate the very helpful comments
and suggestions of two anonymous reviewers, the Associate Editor and the Co-Editor.
Li's research is supported in part by NSFC (Nos. 11571348 and 11771239)
and Tsinghua University Initiative Scientific Research Program (No. 2019Z07L01009).
Zhu's research is supported in part by RGC of Hong Kong (No. 17306818), NSFC (Nos. 11571348, 11690014,
11731015 and 71532013), Seed Fund for Basic Research (Nos. 201611159233 and 201811159049) and Hung Hing Ying
Physical Sciences Research Fund 2017-18.

\appendix
\section{proofs of theorems}

Recall that $J_n(\theta)=\frac{\partial^2F_n(\theta)}{\partial\theta\partial\theta'}$. In what follows, we let
$I_{n}(\theta)=\frac{\partial F_n(\theta)}{\partial\theta}$ and $C$ be a positive generic constant. As for $\theta_0=(\theta_0^{(1)'},\theta_0^{(2)'},\theta_0^{(3)'})'$, we write
$Z=(Z_{(1)}',Z_{(2)}',Z_{(3)}')'$ and $G:=JZ=(G_{(1)}',G_{(2)}',G_{(3)}')'$,
where $Z_{(i)}$ or $G_{(i)}$ is of dimension $d_i$ for $i=1,2,3$.
Meanwhile, we let $J_{(ij)}$, $i,j=1,2,3$, be the corresponding blocks of the matrix $J$, and $\Lambda_{(i)}$, $i=1,2,3$, be the corresponding blocks of the space $\Lambda$. Further, we let $Z_{(2,3)}=(Z_{(2)}',Z_{(3)}')'$, $\Lambda_{(2,3)}=\Lambda_{(2)}\times \Lambda_{(3)}$,
$\lambda^{\Lambda}_{(2,3)}$ be the last $(d_2+d_3)$ components of $\lambda^{\Lambda}$, and
\[\lambda_{(2,3)}^{\Lambda_{(2,3)}}=\arg\inf_{\lambda_{(2,3)}\in\Lambda_{(2,3)}}\left\|Z_{(2,3)}-\lambda_{(2,3)}\right\|^2_{({KJ^{-1}K'})^{-1}}.\]

\subsection*{Proof of Theorem \ref{mainthm}}
(i) By Lemma \ref{l2} and the similar arguments as for Theorem 2.1 in \cite{zhu2011global}, we can show that (i) holds.

(ii) Let
\begin{flalign}\label{argmin}
\theta_{J_n}(Z_n)=\arg\inf_{\theta\in\Theta}\| Z_n-\sqrt{n}(\theta-\theta_0)\|_{J_n} \quad\text{and}\quad \lambda_n^{\Lambda}=\arg\inf_{\lambda\in\Lambda}\left\| Z_n-\lambda\right\|_{J_n},
\end{flalign}
where
\begin{flalign}\label{J_and_Z}
J_n=J_{n}(\theta_0)\quad\text{and}\quad
Z_n=-J_n^{-1}\sqrt{n}I_{n}(\theta_0).
\end{flalign}
By Lemma \ref{l3}(i)-(ii), $J$ is positive definite and $J_{n}-J=o_{p}(1)$. Then, by Lemma 2.3 in \cite{KK2011}, it is not hard to show that $J_{n}$ is a positive definite matrix (a.s.) for sufficiently large $n$.
Since $\Theta$ contains a hypercube, a similar argument as \cite{francq2007quasi} implies that
\begin{equation}\label{lambdan}
\sqrt{n}\big(\theta_{J_n}(Z_n)-\theta_0\big)=\lambda_n^\Lambda
\end{equation}
for sufficiently large $n$. Using Taylor's expansion for a function with right partial derivatives we get
for all $\theta$ and $\theta_0 \in \Theta$,
\begin{flalign}
F_n(\theta)=&F_n(\theta_0)+I_{n}(\theta_0)'(\theta-\theta_0)+\frac{1}{2}(\theta-\theta_0)'J_{n}(\theta_0)(\theta-\theta_0)+R_n(\theta) \nonumber\\ =&F_n(\theta_0)-\frac{1}{2n}Z_n'J_n\sqrt{n}(\theta-\theta_0)-\frac{1}{2n}\sqrt{n}(\theta-\theta_0)'J_nZ_n+\frac{1}{2}(\theta-\theta_0)'J_n(\theta-\theta_0)+R_n(\theta) \nonumber \\
=&F_n(\theta_0)+\frac{1}{2n}\| Z_n-\sqrt{n}(\theta-\theta_0)\|^2_{J_n}-\frac{1}{2n}\|Z_n\|_{J_n}+R_n(\theta), \label{taylor1}
\end{flalign}
where $R_n(\theta)$ is the  remainder. Next, by Lemma \ref{l3}(ii) and the similar arguments as for Theorem 2 in
\cite{francq2007quasi}, we can show that
\begin{enumerate}
	\item[(ai)] $\sqrt{n}\big(\theta_{J_n}(Z_n)-\theta_0\big)=O_p(1)$;
	\item[(aii)] $\|\sqrt{n}(\hat{\theta}_n-\theta_0)\|_{J_n}=O_p(1)$;
	\item[(aiii)] for any sequence $(\tilde{\theta}_n)$ such that $\sqrt{n}(\tilde{\theta}_n-\theta_0)=O_p(1)$,\,\,
	$R_n(\tilde{\theta}_n)=o_p(n^{-1})$;
	\item[(aiv)] $\| Z_n-\sqrt{n}(\hat{\theta}_n-\theta_0)\|^2_{J_n}-\|
	Z_n-\lambda_n^\Lambda\|^2_{J_n}=o_p(1)$;
	\item[(av)] $\sqrt{n}(\hat{\theta}_n-\theta_0)-\lambda_n^\Lambda=o_p(1)$;
    \item[(avi)] $\lambda_n^\Lambda\rightarrow_{\mathcal{L}}\lambda^\Lambda$.
\end{enumerate}

\noindent Now, the conclusion in (ii) holds by (av) and (avi). This completes all of the proofs.
\hfill $\square$

\subsection*{Proof of Theorem \ref{testmain}}
By Lemma \ref{l3}(ii), we have that under $H_{0}$,
\begin{flalign*}
\hat{\Sigma}_n=\Sigma+o_p(1), \,\,\,\hat{J}_n=J+o_p(1),\,\,\,\hat{\Sigma}_{n|3}=\Sigma+o_p(1)\,\,\,\mbox{and}\,\,\,\hat{J}_{n|3}=J+o_p(1).
\end{flalign*}

(i) Note that
$\sqrt{n}(\hat{\theta}_n^{(3)}-\theta_0^{(3)})=\sqrt{n}K_{\alpha}(\hat{\theta}_n-\theta_0)\rightarrow_{\mathcal{L}}K_{\alpha}\lambda^{\Lambda}$ by  Theorem \ref{mainthm}. Thus, it follows by the continuous mapping theorem that $W_n\rightarrow_{\mathcal{L}}W=\lambda^{\Lambda'}\Omega\lambda^{\Lambda}$.

(ii) Rewrite the Lagrangian  as $F_n(\theta)+(K_{\alpha}\theta)'\gamma$ for a Lagrange multiplier $\gamma$. Then, the solutions
$\hat{\theta}_{n|3}$ and $\gamma_{n|3}$ satisfy the saddle-point condition:
\begin{equation*}
I_n(\hat{\theta}_{n|3})+K_{\alpha}'\gamma_{n|3}=0.
\end{equation*}
By Taylor's expansion, it follows that
\begin{flalign*}
0=\sqrt{n}I_n(\theta_0)+\sqrt{n}J_n(\theta^*)(\hat{\theta}_{n|3}-\theta_0)+\sqrt{n}K_{\alpha}'\gamma_{n|3}+o_p(1),
\end{flalign*}
where $\theta^*$ lies between $\hat{\theta}_{n|3}$ and $\theta_0$. Since $J_n(\theta^*)=J+o_{p}(1)$ under $H_{0}$ by Lemma \ref{l3}(ii), the preceding equation entails that
\begin{flalign}\label{taylor2}
\sqrt{n}I_n(\theta_0)+\sqrt{n}J(\hat{\theta}_{n|3}-\theta_0)+\sqrt{n}K_{\alpha}'\gamma_{n|3}=o_p(1).
\end{flalign}
Multiplying both sides of (\ref{taylor2}) by $K_{\alpha}J^{-1}$,
we have
$\sqrt{n}\gamma_{n|3}=-(K_{\alpha}J^{-1}K_{\alpha}')^{-1}K_{\alpha}J^{-1}
[\sqrt{n}I_n(\theta_0)]+o_p(1)$
by noting the restriction $K_{\alpha}(\hat{\theta}_{n|3}-\theta_0)=0$. Thus, since
\begin{equation} \label{gamma}
\sqrt{n}I_{n}(\theta_0)\rightarrow _{\mathcal{L}}
\mathcal{N}(0, \Sigma)
\end{equation}
by Lemma \ref{l1}(ii) and the martingale central limit theorem in \cite{Brown},
we can conclude that
\begin{flalign*}
\sqrt{n}\gamma_{n|3}\rightarrow _{\mathcal{L}} \mathcal{N}(0, \,(K_{\alpha}J^{-1}K_{\alpha}')^{-1}K_{\alpha}J^{-1}\Sigma J^{-1}K_{\alpha}'(K_{\alpha}J^{-1}K_{\alpha}')^{-1}),
\end{flalign*}
which implies that $L_n\rightarrow _{\mathcal{L}}\chi_{d_3}^2$ by the continuous mapping theorem,
and the facts that both $\Sigma$ and $J$ are positive definite by Lemma \ref{l3}(i).

(iii) By Taylor's expansion and the similar proof as for Theorem \ref{mainthm}, we can get
\begin{flalign*}
nF_n(\hat{\theta}_{n})=&nF_n(\theta_0)+nI_n(\theta_0)(\hat{\theta}_{n}-\theta_0)+\frac{n}{2}(\hat{\theta}_{n}-\theta_0)'
J(\hat{\theta}_{n}-\theta_0)+o_p(1)\\=&
nF_n(\theta_0)+\frac{1}{2}(\left\|Z-\lambda^{\Lambda}\right\|_J^2-\left\|Z\right\|_J^2)+o_p(1)
\end{flalign*}
and
\begin{flalign*}
nF_n(\hat{\theta}_{n|3})=&nF_n(\theta_0)+nI_n(\theta_0)(\hat{\theta}_{n|3}-\theta_0)+\frac{n}{2}(\hat{\theta}_{n|3}-\theta_0)'
J(\hat{\theta}_{n|3}-\theta_0)+o_p(1)\\
=&nF_n(\theta_0)+\frac{1}{2}(\big\|Z-\lambda_{|3}^{\Lambda}\big\|_J^2-\left\|Z\right\|_J^2)+o_p(1).
\end{flalign*}
Hence, it follows that $Q_{n}\rightarrow _{\mathcal{L}} Q=\big\|Z-\lambda_{|3}^{\Lambda}\big\|_J^2-\big\|Z-\lambda^{\Lambda}\big\|_J^2$.

Note that
\begin{flalign*}
\big\|Z-\lambda^{\Lambda}\big\|_J^2-\left\|Z\right\|_J^2
=&-G_{(1)}'J_{(11)}^{-1}G_{(1)}-2\lambda_{(2,3)}^{\Lambda'}(KJ^{-1}K')^{-1}Z_{(2,3)}+\lambda_{(2,3)}^{\Lambda'}(KJ^{-1}K')^{-1}\lambda_{(2,3)}^{\Lambda}\\
=&-G_{(1)}'J_{(11)}^{-1}G_{(1)}-2\lambda_{(2,3)}^{\Lambda_{(2,3)}'}(KJ^{-1}K')^{-1}Z_{(2,3)}+\lambda_{(2,3)}^{\Lambda'}(KJ^{-1}K')^{-1}\lambda_{(2,3)}^{\Lambda_{(2,3)}}\\=&
-G_{(1)}'J_{(11)}^{-1}G_{(1)}-\lambda_{(2,3)}^{\Lambda_{(2,3)}'}(KJ^{-1}K')^{-1}\lambda_{(2,3)}^{\Lambda_{(2,3)}}
+2\lambda_{(2,3)}^{\Lambda_{(2,3)}'}(KJ^{-1}K')^{-1}(\lambda_{(2,3)}^{\Lambda_{(2,3)}}-Z_{(2,3)})\\
=&-G_{(1)}'J_{(11)}^{-1}G_{(1)}-\lambda^{\Lambda'}\Xi\lambda^{\Lambda},
\end{flalign*}
where the first equation follows by Lemma \ref{l6}(i)-(ii), the second equation follows by
Lemma \ref{l6}(iii), and
the third equation holds since
$\lambda_{(2,3)}^{\Lambda_{(2,3)}}=K \lambda^{\Lambda}$ and $\lambda_{(2,3)}^{\Lambda_{(2,3)}}$ is orthogonal to $(\lambda_{(2,3)}^{\Lambda_{(2,3)}}-Z_{(2,3)})$.
Using the same technique to $\big\|Z-\lambda_{|3}^{\Lambda}\big\|_J^2-\left\|Z\right\|_J^2$,
we can obtain that $Q=\lambda^{\Lambda'}\Xi\lambda^{\Lambda}-\lambda_{|3}^{\Lambda'}\Xi\lambda_{|3}^{\Lambda}$.
This completes all of the proofs.
\hfill $\square$

\subsection*{Proof of Theorem \ref{one coefficient}}
When $H_0'$ holds and $d_{2}=0$, $\Lambda=R^{d-1}\times[0,\infty)$ and
$K=K_{\alpha}=(0,\cdots,0,1)$. In  this case,
\begin{flalign*}
\lambda^{\Lambda}=ZI(Z_d\geq0)+PZI(Z_d<0)=Z-Z_d^{-}c^{*},
\end{flalign*}
where $P=I_d-J^{-1}K'(KJ^{-1}K')^{-1}K$, $Z_d^{-}=Z_dI(Z_d<0)$,
and the vector $c^{*}$ is the last column of $J^{-1}$ divided by the $(d,d)$ element of $J^{-1}$.

First, since the last element of $c^{*}$ is 1, the $d$th element of $\lambda^{\Lambda}$ is  $\lambda_d^{\Lambda}=Z_dI(Z_d\geq0)$.
Thus, it follows that
\begin{flalign}\label{wu}
W=\lambda^{\Lambda'}\Omega\lambda^{\Lambda}=[\lambda_d^{\Lambda}]^{2}/\mathrm{var}(Z_d)
=Z^2_dI(Z_d\geq0)/\mathrm{var}(Z_d) \sim U^2I(U\geq0),
\end{flalign}
where $U\sim \mathcal{N}(0,1)$. By Theorem \ref{testmain}(i) and (\ref{wu}), we know that $W_{n}$ has the critical region $\{W_{n}>\chi^{2}_{1,1-2\beta}\}$.

Second, $t_{n}$ has the critical region $\{t_{n}>\Phi^{-1}(1-\beta)\}$, since $t_n=\sqrt{W_n}$ has the law of $UI(U\geq0)$.

Third, since $\xi Q=W$ with $\xi=\Omega/\Xi$, we know that  $\xi Q$ has the law of $W$. By Theorem \ref{testmain}(iii), it follows that
$Q_{n}$ has the critical region
$\{\xi_{n} Q_{n}>\chi^{2}_{1,1-2\beta}\}$, where
$\xi_{n}=\hat{\Omega}_n/\hat{\Xi}_n$ is a consistent estimator of $\xi$. This completes all of the proofs.
\hfill $\square$

\subsection*{Proof of Theorem \ref{mainthm2}}
(i) Note that under $\mathbb{P}_{n,h}$, $F_{n}(\theta)=\frac{1}{n}\sum_{t=1}^{n}w_{t,n}\ell_{t,n}(\theta)$, where
$w_{t,n}$ and $\ell_{t,n}(\theta)$ are defined as in (\ref{local}).
First, by Assumption \ref{a.3p} with $\delta=0$ and a similar argument as for Lemma \ref{l1}(i),
it is not hard to show that
\begin{flalign}
E\sup_{\theta\in\Theta}w_{t,n}^2 \ell^2_{t,n}(\theta)<\infty.
\end{flalign}
Next, by Theorem 2.1 in \cite{ling2007double}, $y_{t,n}$ is geometrically ergodic, and hence $\beta$-mixing. By Theorem 3.49 in \cite{white}, it follows that
$\ell_{t,n}(\theta)$ is $\beta$-mixing. Then, by Lemma \ref{lln} with $s=2$, we have
\begin{equation}\label{localconsistency}
\lim\limits_{n\rightarrow\infty}\frac{1}{n}\sum_{t=1}^{n}[w_{t,n} \ell_{t,n}(\theta)-Ew_{t,n} \ell_{t,n}(\theta)]=0 \quad \text{in probability}.
\end{equation}
Furthermore, the stationarity of $w_{t,n}\ell_{t,n}(\theta)$ ensures that
\begin{flalign*}
\frac{1}{n}\sum_{t=1}^{n}Ew_{t,n}\ell_{t,n}(\theta)=Ew_{t,n}\ell_{t,n}(\theta),
\end{flalign*}
and the dominated convergence theorem entails that
\begin{flalign*}
\lim\limits_{n\rightarrow\infty}Ew_{t,n}\ell_{t,n}(\theta)
=E\lim\limits_{n\rightarrow\infty}w_{t,n}\ell_{t,n}(\theta)=Ew_t \ell_t(\theta).
\end{flalign*}
By (\ref{localconsistency}) and the preceding equalities, we know that for any $\theta\in\Theta$,
\begin{flalign}\label{result_pointwise}
F_{n}(\theta)-Ew_t\ell_t(\theta)=o_{p}(1).
\end{flalign}
Third, by a similar argument as for Lemma \ref{l1}(ii), we can prove that
$\sup_{\theta\in\Theta}\left\|I_{n}(\theta)\right\|=O_p(1)$, and hence it follows that
Assumption 3A in \cite{newey1991} holds, and then by (\ref{result_pointwise}) and  Theorem 2.1 in \cite{newey1991},
we have
\begin{flalign}\label{result_uniform}
\sup_{\theta\in\Theta}|F_{n}(\theta)-Ew_t \ell_t(\theta)|=o_p(1).
\end{flalign}
Finally, since $E[w_t \ell_t(\theta)]$  attains the global minimum at $\theta_0$ by Lemma \ref{l2}(i),
(i) holds by (\ref{result_uniform}) and Theorem 4.1.1 in \cite{amemiya1985}.

(ii) Recall that $\theta_n=\theta_0+h/\sqrt{n}$.  Let
\begin{flalign*}
\theta_{J_{n,h}}(Z_{n,h})=\arg\inf_{\theta\in\Theta}\left\|Z_{n,h}-\sqrt{n}(\theta-\theta_n)\right\|_{J_{n,h}}
\quad\mbox{and} \quad
\lambda_{n,h}^{\Lambda}=\arg\inf_{\lambda\in\Lambda}\|Z_{n,h}+h-\lambda\|_{J_{n,h}},
\end{flalign*}
where
\begin{flalign}\label{Jh_Zh}
J_{n,h}=J_{n}(\theta_n)\quad\mbox{and} \quad
Z_{n,h}=-J_{n,h}^{-1}\sqrt{n}I_{n}(\theta_n).
\end{flalign}
By Lemma \ref{l3}(i) and (iii), we can show that $J_{n,h}$ is a positive definite matrix (a.s.) for sufficiently large $n$.
Next, by the similar arguments as for (\ref{lambdan}) and (\ref{taylor1}), we have that
\begin{equation*}
\sqrt{n}\big(\theta_{J_{n,h}}(Z_{n,h})-\theta_n\big)=\lambda_{n,h}^\Lambda-h
\end{equation*}
for sufficiently large $n$, and
\begin{flalign*}  
F_{n}(\theta)=F_{n}(\theta_n)+\frac{1}{2n}\left\|Z_{n,h}-\sqrt{n}(\theta-\theta_n)
\right\|^2_{Z_n,h}-\frac{1}{2n}\|Z_{n,h}\|_{J_{n,h}}+R_{n}(\theta),
\end{flalign*}
where $R_{n}(\theta)$ is the remainder.

Furthermore, by Lemma \ref{l3}(iii) and the similar arguments as for (ai)-(avi), we can show that
\begin{enumerate}
	\item[(bi)] $\sqrt{n}\big(\theta_{J_{n,h}}(Z_{n,h})-\theta_n\big)=O_p(1)$;
	\item[(bii)] $\|\sqrt{n}(\hat{\theta}_{n,h}-\theta_n)\|_{J_{n,h}}=O_p(1)$;
	\item[(biii)] for any sequence $(\tilde{\theta}_n)$ such that $\sqrt{n}(\tilde{\theta}_n-\theta_n)=O_p(1)$,\,\,
    $R_{n}(\tilde{\theta}_n)=o_p(n^{-1})$;
	\item[(biv)] $\| Z_{n,h}-\sqrt{n}(\hat{\theta}_{n,h}-\theta_n)\|^2_{J_{n,h}}-\|
    Z_{n,h}-(\lambda_{n,h}^\Lambda-h)\|^2_{J_{n,h}}=o_p(1)$;
	\item[(bv)] $\sqrt{n}(\hat{\theta}_{n,h}-\theta_n)-(\lambda_{n,h}^\Lambda-h)=o_p(1)$;
	\item[(bvi)] $\lambda_{n,h}^{\Lambda}\rightarrow_{\mathcal{L}}\lambda^{\Lambda}(h)$.
\end{enumerate}
\noindent Now, (ii) holds by (bv) and (bvi). This completes the proof.
\hfill $\square$



\subsection*{Proof of Theorem \ref{locallimit}}
The proof is given in the supplementary material (\cite{jlz19}).

\subsection*{Proof of Theorem \ref{localpower}}
We first consider the Wald test.
When $d_2=0$ and $d_3=1$,  $\Omega=1/\mathrm{var}(Z_d)=1/\sigma_d^2$ by (\ref{omega}), and
$\lambda_d^{\Lambda}(h)=(Z_d+h_d)I(Z_d+h_d>0)$ by the similar arguments as for (\ref{wu}).
Hence, it follows that
\begin{flalign}\label{local_Wald}
W(h)=\lambda_d^{\Lambda}(h)^2/\mathrm{var}(Z_d)=\left(\frac{Z_d}{\sigma_d}+\frac{h_d}{\sigma_d}\right)^2I(Z_d+h_d>0)
\sim (U+h^{*})^2I({U+h^{*}>0})
\end{flalign}
for $U\sim \mathcal{N}(0,1)$ and $h^{*}=h_d/\sigma_d$. By (\ref{local_Wald}), we know that the result for the Wald test holds.
Furthermore, since $\xi Q(h)= W(h)$, it follows that the result for the QLR test holds.

Next, we consider the $t$-type test. Since $\{(U+h^*)>c_1\}\subset\{(U+h^*)>0\}$, we have
\begin{flalign}\label{local_t}
\mathbb{P}\{(U+h^*)I({U+h^*>0})>c_1\}=\mathbb{P}\{(U+h^*)>c_1\}.
\end{flalign}
By noting that $W_n=t_n^2$ and (\ref{local_Wald})-(\ref{local_t}), we can show that the result for the $t$-test holds.

Third, we consider the LM test. Since $L(h)\sim \chi^2_{1}(h'\Omega h)\sim (U+h^*)^2$ and
\[
\mathbb{P}\{(U+h^*)^2>c_2^2\}=
\mathbb{P}\{(U+h^*)>c_2\}+\mathbb{P}\{(U+h^*)<-c_2\},
\]
it follows that the result for the LM test holds.

Finally, by the similar arguments as in \cite{francq2009testing}, we can prove that the asymptotic local power of the $t$-type, Wald and QLR tests is higher than
that of the LM test. The proof is completed.   \hfill $\square$


\subsection*{Proof of Theorem \ref{jointdistribution}}
The proof  is given in the supplementary material (\cite{jlz19}).
\hfill $\square$

	\section{LEMMAS}\label{sec:lemmas}
This appendix provides five useful lemmas, and their proofs can be found in the supplementary material (\cite{jlz19}).

	\begin{lemma}\label{l1}
		Suppose that Assumptions \ref{a.1}-\ref{a.3} hold. Then,

		$\mathrm{(i)}$  ${\displaystyle E\sup_{\theta\in\Theta}|w_t\ell_t(\theta)|<\infty}$;
			
       $\mathrm{(ii)}$ ${\displaystyle E\sup_{\theta\in\Theta}\left\| w_t\frac{\partial \ell_t(\theta)}{\partial\theta}\right\|<\infty}$;
			
       $\mathrm{(iii)}$ ${\displaystyle E\sup_{\theta\in\Theta}\left\| w_t\frac{\partial^2 \ell_t(\theta)}{\partial\theta\partial\theta'}\right\|<\infty}$.
	\end{lemma}


	\begin{lemma}\label{l2}
		For any $\theta^{*}\in\Theta$, let $B_{\eta}(\theta^{*})=\{\theta\in\Theta: \|\theta-\theta^{*}\|<\eta\}$ be an
open neighborhood of $\theta^{*}$ with radius $\eta>0$. Suppose that Assumptions \ref{a.1}-\ref{a.3} hold. Then,
		
$\mathrm{(i)}$  $E[w_t\ell_t(\theta)]$ has a unique minimum at $\theta_0$.

$\mathrm{(ii)}$ ${\displaystyle E\left[\sup_{\theta\in B_{\eta}(\theta^{*})} w_{t}|\ell_{t}(\theta)-\ell_{t}(\theta^{*})|\right]\to 0}$ as $\eta\to0$.
	\end{lemma}


		\begin{lemma}\label{lln}
		Let $\{X_{ni}:1\leq i\leq k_n, n=1,2,\cdots\}$ be a mean-zero triangular array of $\beta$-mixing sequences that are $L^s$-bounded for some $s>1$, and $\mathcal{F}_{ni}=\sigma(X_{n1},\cdots,X_{ni})$ for $1\leq i\leq k_n$. Then $\{X_{ni},\mathcal{F}_{ni}\}$ is a uniformly integrable $L^1$-mixingale. Furthermore,
		$E\big|\frac{1}{k_n}\sum_{i=1}^{k_n}X_{ni}\big|\rightarrow0$ as $n\rightarrow\infty$, and hence $\frac{1}{k_n}\sum_{i=1}^{k_n}X_{ni}\rightarrow0$ in probability as $n\rightarrow\infty$.
	\end{lemma}


	\begin{lemma}\label{l3}
Let $\theta_n^*$ be a sequence such that $\theta_n^*\rightarrow\theta_0$ in probability. Suppose that Assumptions \ref{a.1}-\ref{a.4} hold. Then,
		
$\mathrm{(i)}$ both $J$ and $\Sigma$ are finite, non-singular and  positive definite matrices;

$\mathrm{(ii)}$ $\left\|J_n(\theta_n^*)-J\right\|=o_p(1)$, $\left\|\Sigma_n(\theta_n^*)-\Sigma\right\|=o_p(1)$, $\left\|D_n(\theta_n^*)-D\right\|=o_p(1)$,   and $Z_n\rightarrow_\mathcal{L}Z$, where  $Z_n$ is defined in (\ref{J_and_Z});

\noindent Furthermore, if Assumption \ref{a.3} is replaced by Assumption \ref{a.3p} with $\delta=2$, under $\mathbb{P}_{n,h}$,

$\mathrm{(iii)}$ $\left\|J_{n}(\theta_n^*)-J\right\|=o_p(1)$, $\left\|\Sigma_{n}(\theta_n^*)-\Sigma\right\|=o_p(1)$, $\left\|D_{n}(\theta_n^*)-D\right\|=o_p(1)$,   and $Z_{n,h}\rightarrow_\mathcal{L}Z$, where $Z_{n,h}$ is defined in (\ref{Jh_Zh}).
	\end{lemma}

	\begin{lemma}\label{l6}
		Suppose that  Assumptions \ref{a.1}-\ref{a.4} hold. Then,
		\begin{enumerate}
			\item[$\mathrm{(i)}$] $Z'JZ=Z_{(2,3)}'(KJ^{-1}K)^{-1}Z_{(2,3)}+G_{(1)}'J_{(11)}^{-1}G_{(1)}$;
			\item [$\mathrm{(ii)}$] $\big\|Z-\lambda^{\Lambda}\big\|^2_J=\big\|Z_{(2,3)}-\lambda_{(2,3)}^{\Lambda}\big\|^2_{({KJ^{-1}K'})^{-1}}
			=\big\|Z_{(2,3)}-\lambda_{(2,3)}^{\Lambda_{(2,3)}}\big\|^2_{({KJ^{-1}K'})^{-1}}$;
			\item[$\mathrm{(iii)}$] $\lambda^{\Lambda}_{(2,3)}=\lambda_{(2,3)}^{\Lambda_{(2,3)}}$.
		\end{enumerate}
	\end{lemma}

\end{document}